\documentclass[useAMS,usenatbib]{mn2e}

\usepackage{graphicx}
\usepackage{scalefnt}
\usepackage{amssymb}

\title[Close-encounters in a group environment]{Simulating the evolution of disc galaxies in a group environment. 
II. The influence of close-encounters between galaxies}
\author[\'A. Villalobos, G. De Lucia, and G. Murante]{
\'A. Villalobos$^{1}$\thanks{villalobos@oats.inaf.it}, G. De Lucia$^{1}$, and G. Murante$^{1,2}$
\\
$^{1}$INAF - Astronomical Observatory of Trieste, via G.B. Tiepolo 11, I-34143 Trieste, Italy\\
$^{2}$INAF - Istituto Nazionale di Astrofisica - Osservatorio Astronomico di Torino, Str. 
Osservatorio 25, I-10025, Pino Torinese, Torino, Italy}

\begin{document}

\date{Accepted --- . Received --- ; in original form ---}

\pagerange{\pageref{firstpage}--\pageref{lastpage}} \pubyear{---}

\maketitle

\label{firstpage}

\begin{abstract}
We study the evolution of disc galaxies in group environments under
the effect of both the global tidal field and close-encounters between 
galaxies, using controlled $N$-body simulations of isolated mergers.
We find that close-range encounters between galaxies are less frequent 
and less damaging to disc galaxies than originally expected, since they 
mostly occur when group members have lost a significant fraction of 
their initial mass to tidal stripping. We also find that group members 
mostly affect disc galaxies \emph{indirectly} by modifying their common 
global tidal field. Different initial orbital parameters of group 
members introduce a significant ``scatter'' in the evolution of general 
properties of disc galaxies around a ``median'' evolution that is 
similar to when only the effect of the global tidal field is included. 
Close-encounters introduce a high variability in the properties of disc 
galaxies, even \emph{slowing} their evolution in some cases, and could 
wash out correlations between galaxy properties and the group total 
mass. The combined effect of the global tidal field and 
close-encounters appears to be inefficient at forming/enhancing central 
stellar bulges. This implies that bulges of S0 galaxies should be 
mostly composed by young stars, which is consistent with recent 
observations.

\end{abstract}


\section{Introduction}

Galaxies experience different environments during their lifetimes, 
inhabiting regions of the Universe ranging in density from ``average'' 
to medium and high in environments such as groups and clusters of 
galaxies \citep[e.g.,][]{lacey1994,zhao2003,berrier2009,mcgee2009}.

In these regions galaxies evolve under the influence of a combination of 
environmental and internal processes. Galaxies infalling onto a halo 
experience the effects of the global tidal field of their host
\citep[e.g.][]{gnedin2003} which can vary gradually in intensity along 
their orbits or more abruptly if the halo grows rapidly or the galaxy's 
stellar mass grows significantly after accretion 
\citep[e.g][]{neistein2011}. Galaxies can also experience gravitational 
interactions with other halo galaxies or substructures in the form of 
direct mergers or close-encounters with low/high relative velocity 
\citep{moore1996,mastropietro2005}. For galaxies in dense environments, 
both the cold and hot reserves of gas can be removed hydrodynamically by 
the hotter intra-group or intra-cluster medium via processes such as 
ram-pressure stripping \citep{gunn1972} and starvation 
\citep{larson1980,balogh2009}, which can limit severely the capacity of 
galaxies to continue forming stars after accretion. Internal galactic 
processes, such as gas cooling, star formation, supernovae feedback, 
active nuclei feedback, formation of central bars also affect the 
evolution of galaxies and their contributions can be triggered/affected 
by environmental factors.

Environmental effects are believed to play a crucial role in the 
evolution of galaxies in dense regions of the Universe. However, the 
precise mechanisms through which they affect galaxies remain unclear 
\citep[see the review by][]{weinmann2011-2}. They are thought to be the 
culprits behind the lower star formation, lower fraction of disc 
morphologies and redder colours observed in galaxies belonging to groups 
and clusters in comparison to regions of lower density 
\citep[e.g.,][]{dressler1997,lewis2002,gomez2003,balogh2004,weinmann2006,mcgee2008}.

Recent studies, based on cosmological simulations combined with 
semi-analytic models, claim that a significant fraction of galaxies that 
currently reside in massive clusters have been accreted as part of galaxy 
groups \citep[][but see also \citeauthor{berrier2009}~\citeyear{berrier2009}]{mcgee2009,delucia2012}.
These results emphasise the concept of galactic ``pre-processing'' 
\citep{zabludoff1998,lisker2013}. 
In this scenario, a considerable fraction of galaxies observed at 
present time in clusters, have had most of their properties shaped 
within group environments. This highlights the importance of 
understanding in detail the role of the environment in the evolution of 
galaxies within groups.

However, theoretical studies on environmental effects within galaxy 
groups remain surprisingly few, while most of the theoretical effort 
has been largely focused on galaxies residing in either massive clusters 
\citep[e.g.][]{moore1999,gnedin2003,mastropietro2005} or in Milky Way-like haloes
\citep[e.g.][]{mayer2001,mayer2001-2,klimentowski2009,kazantzidis2011}.
Most of the recent studies on the evolution of galaxies in group-like 
environments by means of observations or numerical simulations
\citep[][]{feldmann2011,tonnesen2012,bahe2013,bekki2013,taranu2013,vija2013,wetzel2013,ziparo2014} 
have concentrated on the formation of elliptical galaxies, the effect 
of ram-pressure stripping on galaxies by the intergalactic medium, and 
on the efficiency of galaxies to form stars while they inhabit a group. 
Interestingly, little attention has been paid to understand better the 
relative contribution of different environmental 
(and environmentally-triggered) processes (e.g. driven by the global 
tidal field, close-encounters between galaxies, rapid halo growth) to 
the overall evolution of galaxies in groups. This is a fundamental step 
in order to comprehend how galaxies currently residing in massive 
clusters have been ``pre-processed''. 

Close-encounters between galaxies are considered one of the main 
contributors to their evolution. Within groups, these encounters take 
place at lower relative velocity with respect to interactions within 
more massive clusters. Given the long duration of the gravitational 
perturbation, galaxy close-encounters in groups can induce significant 
changes in the structure of galaxies.

\citet[][]{bekki2011} investigate the combined effect of the global 
tidal field of groups and repetitive close-encounters on the evolution 
of spiral galaxies, using chemodynamical simulations. They find that 
the combined tidal interaction can have a significant impact on the 
morphology of galaxies, inducing series of bursts of star formation that 
increase the stellar mass in the central bulges. These results suggest 
that groups are suitable environments where S0 galaxies can be formed, 
in line with implications from observational evidence \citep{wilman2009,just2010}.

As mentioned above, however, the relative contribution of the global 
tidal field and close-encounters to the general evolution of galaxies in 
groups remains unclear. Group members (except the galaxy under study) are 
usually modelled as point mass objects. In addition, the use of standard 
Schechter functions for the stellar mass distribution of group members 
does not allow the evolution that group members experience in terms of 
their orbits, structure and mass content to be followed consistently, 
starting from stellar masses consistent with recent observations. 
Moreover, by adopting a significantly higher number of group members in 
comparison to observations \citep[as done for example in][]{bekki2011}, 
the contribution of close-encounters to galaxy evolution could be 
largely overestimated. All these factors are potentially relevant for 
the way galaxies interact with each other, and perhaps more importantly, 
for how galaxies might also affect the environment they inhabit. In 
groups, the co-evolution between a single galaxy and its environment 
should be more significant than in clusters, because of its larger 
contribution to the total mass of the environment \citep[see][]{aceves2013}.

In the previous paper of this series \citep{villalobos2012}, we explore 
the evolution of disc galaxies within the global tidal field of a galaxy 
group by means of $N$-body simulations of isolated mergers, covering an 
ample parameter space of different orbits, disc inclinations, 
galaxy-to-group mass ratios, and presence of a central bulge. We find that 
the galaxy-to-group mass ratio and the initial inclination of disc 
galaxies at the time of accretion play a fundamental role in the evolution 
of galaxies within a group \citep[see also][]{bekki2013}. Specifically, we 
find that disc galaxies start suffering significant evolution due to the 
global tidal field only after the mean density of the group (within the 
orbit of the galaxy) exceeds 0.3--1 times the disc galaxy central density. 
Different inclinations of disc galaxies at the time of accretion cause 
them to experience significantly different structural evolution. In 
particular, retrograde infalls (inclination of 180$\degr$ with respect of 
their orbital plane) allow accreted galaxies to retain their initial disc 
structure for a significantly longer time in comparison to prograde infalls 
(inclination of 0$\degr$). Additionally, we find that the global tidal 
field of a group environment is not efficient at either inducing the 
formation of central bulges in stellar discs or enhancing existing ones at 
the time of accretion. This result suggests that the global tidal field 
alone cannot explain the formation of S0 galaxies in groups, and that 
additional processes (e.g., close-encounters, internal processes) are 
required to explain the formation of additional bulge stars. Finally, we 
find that more massive galaxies suffer more tidal stripping due to the fact 
that the stronger dynamical friction acting on them drags them rapidly to the 
densest region of the group, where they are exposed to stronger tidal forces.

In this paper, we present the results of controlled $N$-body simulations 
of infalling disc galaxies as they are accreted onto a group-size halo that 
contains a central galaxy and a population of satellite galaxies. We explore 
different orbital parameters (consistent with cosmological simulations) for 
the infalling disc galaxy and various spatial, velocity and mass distributions 
for the population of satellite galaxies. Our main goal is to determine the 
contribution of low-velocity close-encounters to the evolution of galaxies 
within a group environment. Our approach is to compare simulations of the 
combined influence of global tidal field and close-encounters to simulations 
where only the influence of the global tidal field is included. 

As in the first paper of this series, our group environments are built from 
idealised initial conditions and do not account for hierarchical growth. 
In particular, our simulations do not consider cases where galaxies are 
accreted and evolve as part of ``sub-haloes'' within a group 
halo\footnote{\citet{delucia2012} estimate that about half of low 
($9 < \log[M_*/M_{\sun}] < 10$) and intermediate ($10 < \log[M_*/M_{\sun}] < 11$) 
mass group members are accreted onto their final group when they are satellite 
galaxies. The remaining half being accreted while being central galaxies or 
isolated ones.}. Note also that our simulations do not include gaseous 
components and star formation, since we focus on the effect of 
close-encounters (and the group tidal field) on the stellar content of 
galaxies that is \emph{already present} at the time they are accreted onto a 
group environment. We plan to explore the effect of internal processes and 
the evolution of star formation in galaxies within groups in the next paper 
of this series.

The layout of this paper is as follows: 
Section~\ref{sec-setup} describes the set-up of our experiments; 
Section~\ref{sec-descrip} describes our findings; 
in Section~\ref{sec-discussion} we discuss our results and in 
Section~\ref{sec-conclusions} we give our conclusions.

\section{Set-up of numerical experiments}
\label{sec-setup}

We have carried out 113 simulations of isolated mergers between a disc galaxy and
a group halo containing a population of galaxies. Our basic setup is to release a
single disc galaxy at a time, on a bound orbit, from the virial radius of the
larger halo, exploring its evolution for 12~Gyr. As the galaxy infalls toward the
central region of the halo, dragged by dynamical friction, it suffers the effects
of both the global tidal field of the group, and also of close-encounters with
other galaxies within the group. Each simulation generates 192 snapshots which
allow us to follow accurately the orbits, mass stripping and overall evolution
of each galaxy.

\subsection{Initialisation of multi-component systems}
The group halo in our simulations is modelled as a $N$-body self-consistent
multi-component system composed by a DM halo that follows a NFW density profile
\citep{navarro1997} and a central stellar spheroid following a Hernquist density
profile \citep{hernquist1990}. Our main
haloes have a total mass of 10$^{13} M_{\sun}$, following the the mass range
reported by \citet{mcgee2009} and \citet{delucia2012}, where significant
environmental effects on galaxies must take place \citep[see also][]{berrier2009}.
Additionally, a population of galaxies are embedded in the group halo which are
also composed by both a NFW halo and a central stellar component following a
Hernquist profile. In the rest of the paper, we refer to these galaxies as
``group members'' or ``group population''. Each disc galaxy is modelled as a
multi-component system with a NFW DM halo, a stellar disc following an
exponential radial density profile and a \emph{sech$^2$} vertical density
profile. In some experiments, we also include a Hernquist central stellar bulge. A
detailed description of the procedure followed to initialise our systems can be
found in \citet{villalobos-helmi2008}.

We have carried out preliminary tests to ensure the stability of each system and
lack of evolution due to either particle or time resolution.

\subsection{Number of group members}
The number of group members in our simulations has been chosen to be consistent
with observations of the average number of satellite galaxies found in haloes
in the mass range $13.1 < \log M_{\rm halo} < 13.3$
and with $^{0.1}M_{\rm r}-5 \log h < -19$, respectively, from the SDSS-DR4 ($z <$0.2)
\citep{yang2008}. The most likely number of satellite members within these limits
is found to be 3.  We have, however, chosen to consider 4 satellite galaxies in
our simulations in order to maximise the number of close-encounters and their
effect on the disc galaxy. We have also explored the effect of 8 satellite
galaxies in our simulations.

\begin{table}
\centering
\begin{minipage}{80mm}
 \caption{Properties of the group environment.}
 \label{group-param}
 \begin{tabular}{@{}llr@{}}
  \hline
  DM Halo            &                      &               \\
  \hline
  Virial mass        & $9.9 \times 10^{12}$ & ($M_{\sun}$)  \\
  Virial radius      & 329.19               & (kpc)         \\
  Concentration      & 4.87                 &               \\
  Circular velocity  & 360.07               & (km s$^{-1}$) \\
  Number particles   & $1.1 \times 10^{6}$  &               \\
  Softening          & 0.32                 & (kpc)         \\
  \hline
  Stellar spheroid   &                      &               \\
  \hline
  Mass               &  $10^{11}$           & ($M_{\sun}$)  \\
  Scale radius       &  1.91                & (kpc)         \\
  Number particles   &  $5 \times 10^{5}$   &               \\
  Softening          &  0.06                & (kpc)         \\
  \hline
 \end{tabular}
\end{minipage}
\end{table}

\begin{table}
\centering
\begin{minipage}{80mm} 
 \caption{Properties of the disc galaxy.}
 \label{disc-galaxies-param}
 \begin{tabular}{@{}llr@{}}
  \hline
  Labels             & \multicolumn{2}{c}{\tiny{REF,BUL$^{(1)}$,CIR,RAD,RET}} \\
  \hline
  DM Halo            &                         &               \\
  \hline
  Virial mass        & $10^{12}$               & ($M_{\sun}$)  \\
  Virial radius      & 258.91                  & (kpc)         \\
  Concentration      & 13.12                   &               \\
  Circular velocity  & 129.17                  & (km s$^{-1}$) \\
  Number particles   & $5 \times 10^{5}$       &               \\
  Softening          & 0.35                    & (kpc)         \\
  \hline
  Stellar disc       &                         &               \\
  \hline
  Disc mass          & $2.8 \times 10^{10}$    & ($M_{\sun}$)  \\
  Scale-length       & 3.5                     & (kpc)         \\
  Scale-height       & 0.35                    & (kpc)         \\
  $Q$                & 2                       &               \\
  Number particles   & $10^{5}$                &               \\
  Softening          & 0.05                    & (kpc)         \\
  \hline
 \end{tabular}
\\(1): A stellar bulge is added at the centre of the disc, with 
$M_{\rm bulge}$=$0.3 M_{\rm disc}$ and $a_{\rm bulge}$=$0.2 R_{\rm D}$.
\end{minipage}
\end{table}

\begin{table}
\centering
\begin{minipage}{80mm}
 \caption{Orbital parameters of the disc galaxy.}
 \label{list-exper}
 \begin{tabular}{@{}lcccc@{}}
  \hline
  Label       & Orbit       & $\theta$    & $(V_r,V_{\phi})$  & $e$      \\
              & $^{(1)}$    & $^{(2)}$    & $^{(3)}$          & $^{(4)}$ \\
  \hline
  \hline REF  & Prograde    & 0$\degr$    & (0.9,0.6)         & 0.86     \\
  \hline BUL  & Prograde    & 0$\degr$    & (0.9,0.6)         & 0.86     \\
  \hline CIR  & Prograde    & 0$\degr$    & (0.6,1.1)         & 0.6      \\
  \hline RAD  & Prograde    & 0$\degr$    & (1.2,0.3)         & 0.97     \\
  \hline RET  & Retrograde  & 180$\degr$  & (0.9,0.6)         & 0.86     \\
  \hline 
 \end{tabular}
\\REF: Reference experiment. Most likely orbital infall. BUL: Stellar bulge 
added to disc in REF experiment. CIR: More ``circular'' orbital infall. RAD: 
More ``radial'' orbital infall. RET: Retrograde orbital infall.
(1): Direction of the orbital infall with respect to the disc rotation. 
(2): Initial angle between the orbital and intrinsic angular momentum of the disc.
(3): Radial and tangential components of the initial velocity of the disc, in units 
of the virial circular velocity of the group. 
(4): Initial orbital eccentricity.
\end{minipage}
\end{table}

\begin{table}
\centering
\begin{minipage}{80mm}
 \caption{Properties of the populations of group members.}
 \label{group-members-props}
 \begin{tabular}{@{}lcccc@{}}
  \hline
  Label               & $M_{\rm DM}$            & $R_{\rm vir}$ & $M_{\rm stars}$         & $a_{\rm stars}$  \\
                      & (10$^{10}$ $M_{\odot}$) & (kpc)         & (10$^{10}$ $M_{\odot}$) & (kpc)            \\
  \hline
  \hline 4g1tBA       & 250                     & 351.39        & 7                       & 1.834            \\
                      & 100                     & 258.91        & 2.8                     & 1.416            \\
                      & 22.6                    & 157.71        & 0.634                   & 0.932            \\
                      & 6.04                    & 101.58        & 0.169                   & 0.642            \\
  \hline 4g1tLO       & 146                     & 293.58        & 4.08                    & 1.575            \\
                      & 82.4                    & 242.72        & 2.31                    & 1.341            \\
                      & 77.0                    & 237.27        & 2.16                    & 1.316            \\
                      & 73.0                    & 233.17        & 2.04                    & 1.297            \\
  \hline 4g1tHI       & 144                     & 292.37        & 4.04                    & 1.571            \\
                      & 119                     & 274.06        & 3.32                    & 1.486            \\
                      & 107                     & 264.65        & 2.99                    & 1.443            \\
                      & 7.25                    & 107.96        & 0.203                   & 0.676            \\
  \hline 4g2tBA       & 289                     & 368.79        & 8.12                    & 1.912            \\
                      & 228                     & 340.77        & 6.38                    & 1.787            \\
                      & 137                     & 287.55        & 3.84                    & 1.548            \\
                      & 97.5                    & 256.73        & 2.73                    & 1.406            \\
  \hline 8g1tBA       & 79.6                    & 239.95        & 2.23                    & 1.329            \\
                      & 70.7                    & 230.65        & 1.98                    & 1.285            \\
                      & 59.3                    & 217.52        & 1.66                    & 1.223            \\
                      & 49.6                    & 204.95        & 1.39                    & 1.163            \\
                      & 33.2                    & 179.28        & 0.930                   & 1.038            \\
                      & 31.5                    & 176.16        & 0.882                   & 1.023            \\
                      & 26.9                    & 167.13        & 0.754                   & 0.979            \\
                      & 23.5                    & 159.77        & 0.657                   & 0.942            \\
  \hline
 \end{tabular}
\\Labels follow a nomenclature that indicates the number of group members (4;8), 
the relative total (DM+stars) mass enclosed adding all group members (1;2) and 
the relative distribution of stellar mass in group members with respect to the 
stellar disc. This can be either less (LO) or more massive (HI), or 
balanced (BA). $M_{\rm DM}$, $M_{\rm stars}$, $R_{\rm vir}$ and $a_{\rm stars}$ 
indicate the DM mass, stellar mass, virial radius and Hernquist scalelength of 
the stellar component for each group member. DM haloes and stellar spheroids are 
modelled with 10$^5$ and 10$^4$ particles, 
respectively. The actual DM mass with which each group member is initialised 
corresponds to \emph{half} the values listed (see Section \ref{sec-setup}). 
\end{minipage}
\end{table}

\subsection{Stellar and DM mass content of group members and disc galaxies}
The initial stellar mass of each group member has been randomly drawn from the
conditional stellar mass function of satellite galaxies (average number of
galaxies as a function of galaxy stellar mass in a DM halo of a given mass)
obtained from SDSS-DR4 observations \citep{yang2009}. This corresponds to a
modified Schechter function:
\begin{equation} \label{csmf}
\Phi_{\rm sat}(M_*|M_h) = \phi^*_s \left(\frac{M_*}{M_{*,s}}\right)^{(\alpha^*_s + 1)} \exp \left[ - \left(\frac{M_*}{M_{*,s}}\right)^2 \right],
\end{equation}
that describes the contribution of satellite galaxies to halo masses within
$12.9 < \log M_{\rm halo} < 13.2$ with $\phi^*_s$=1.96, $\alpha^*_s$=$-$1.15 and
$\log M_{*,s}$=10.67.

The stellar mass of the disc galaxy is chosen to be comparable to that of the
Milky Way's stellar disc.

For simplicity, the same DM-to-stellar mass ratio is used for the disc galaxy and 
group members, $\sim$40:1. However, in our simulations group members are 
initialised to resemble a galaxy population on their way to ``relax'' within the 
group halo, as opposed to an ``infalling'' galaxy population. To this aim, haloes 
of group members are initialised with \emph{half} the DM mass corresponding to 
their stellar mass (i.e. with an effective DM-to-stellar mass ratio $\sim$20:1), 
having effective radii that are also half their respective virial radii. This is 
based on our previous study of the effect of the global tidal field of groups on 
galaxies \citetext{\citeauthor{villalobos2012}~\citeyear{villalobos2012}, 
see also \citeauthor{chang2013}~\citeyear{chang2013}}, which shows that infalling 
galaxies lose approximately half of their DM content after the first pericentric 
passage about the group centre.  

Our tests show that larger and smaller DM-to-stellar mass ratios for group 
members, compared to the chosen one, lead to fewer and weaker close-encounters, 
respectively. More massive group members are affected more efficiently by dynamical 
friction, which causes them to sink earlier than the disc galaxy. This leaves 
little time for close-encounters to take place. On the other hand, since less 
massive group members are less affected by dynamical friction they have longer 
infall times with respect to the disc galaxy. Even though this leaves plenty of 
time for close-encounters, they are generally weak and cause little damage to the 
disc galaxy. 

\subsection{Initial orbital parameters of group members and disc galaxies}
\label{orb-params-members}
The orbital positions and velocities of group members are assigned after randomly
replacing particles of the group DM halo with the multi-component $N$-body 
systems described above. As a first approximation, this is consistent with 
observations \citep[e.g.,][]{lin2004,biviano2010} showing that the projected 
number density of galaxies in clusters can be described with NFW density 
profiles.  

Following this procedure, the spatial distribution of group members populates 
randomly the volume of the group halo. However, we apply a set of conditions to 
the random choice of initial positions (and respective velocities) for group 
members in order to avoid that they are initialised too close to each other, to 
the disc galaxy, to the centre of the group, or outside the virial radius of the 
group. Specifically, the initial orbital position $R_{\rm i}$ (and respective 
velocity) for the 
$i$th-group member is assigned only if \emph{all} of the following conditions are 
simultaneously satisfied: 
(i)  $\Delta R_{\rm i,j} \geq (R^{\rm eff}_{\rm i} + R^{\rm eff}_{\rm j})$, 
(ii) $\Delta R_{\rm i,disc} \geq (R^{\rm eff}_{\rm i} + R^{\rm vir}_{\rm disc})$,
(iii) $R_{\rm i} \geq R^{\rm eff}_{\rm i}$, and 
(iv) $(R_{\rm i} + R^{\rm eff}_{\rm i}) \leq R^{\rm vir}_{\rm group})$,
where $\Delta R_{\rm i,j}$ ($\Delta R_{\rm i,disc}$) is the separation between 
the centres of mass of the $i$-th and $j$-th group member (disc galaxy), and 
$R^{\rm eff}_{\rm i}$ is the effective radius of the $i$-th group member (i.e. 
half its virial radius $R^{\rm vir}_{\rm i}$).

Before placing a group member at its assigned position, a spherical volume 
containing an equivalent amount of DM mass is \emph{locally} removed from the 
group halo, in order to keep constant the total amount of mass of the system 
within the virial radius of the group. Our tests show that this removal of DM 
mass from the group halo does not cause significant or long-lasting perturbations 
to its radial density profile.

As in \citet{villalobos2012}, the initial orbital parameters of the disc galaxy 
are chosen to be consistent with the distribution of orbital parameters of 
infalling substructures at the time they cross the virial radius of their parent 
halo, as extracted from cosmological simulations \citep[e.g.][]{benson2005}. 

\subsection{Parameter space coverage}

Tables~\ref{group-param} and \ref{disc-galaxies-param} show a list of the 
structural, kinematical and numerical properties of the group halo and disc 
galaxy used in our simulations. Note that with respect to our previous 
simulations on the effect of the global tidal field of groups on galaxies, we 
have chosen a comparatively smaller group halo for the disc galaxy under study.
This deliberate choice has been taken in order to maximise the frequency and/or 
effectivity of close-encounters within the group halo, and their 
effect on the disc galaxy.   

Our experiments cover a parameter space related to several properties of the disc 
galaxy and of the population of group members that are potentially relevant for 
the evolution of the disc galaxy under the combined effect of the global tidal 
field and close-encounters. 

For the disc galaxy, we explore both prograde and retrograde infalling orbits 
(with respect to its direction of rotation), the most likely and extreme infall 
eccentricities (according to cosmological simulations), and the effect of a 
central stellar bulge (see Table~\ref{list-exper}).  

Regarding the population of group members, we probe the effect of different 
number of members, total mass, and mass distribution (see 
Table~\ref{group-members-props}).

In our simulations, we explore combinations of one aspect of the parameter space 
related to the disc galaxy and one related to the population of group members (in 
all experiments the same group halo is used). Additionally, for each combination 
we explore 12 different initial orbital distributions (i.e. initial positions and 
velocities) for the group members, satisfying the conditions described in 
Section~\ref{orb-params-members}. We also perform ``control'' simulations only 
with the group halo and disc galaxy (group members are not included) aiming to 
assess the net effect of close-encounters on the evolution of the infalling disc 
galaxy. Following the labels introduced in Tables~\ref{list-exper} and 
\ref{group-members-props}, we refer to a given experiment using the following 
notation: 'ddd' for the ``control'' simulation exploring a parameter of the disc 
galaxy, 'ddd.gggggg' for a combination of parameters of the disc galaxy and the 
group population (and regarding all 12 initial orbital distributions explored for 
group members). When a specific initial orbital distribution of group members 
needs to be referenced, a postfix 'n' is added as 'ddd.gggggg.n'. E.g., the 
experiment REF.4g1tBA refers to the disc galaxy infalling with the most likely 
orbital eccentricity onto a group with a population of group members composed by 
4 satellite galaxies with a relative total mass equals to 1 and a ``balanced'' 
stellar mass distribution with respect to the stellar disc.

It is important to highlight that the total mass in \emph{all} experiments is 
kept approximately constant ($\approx$10$^{13}M_{\odot}$) to facilitate the 
comparison with the ``control'' simulations.

\begin{figure*}
\begin{center}
\includegraphics[width=120mm]{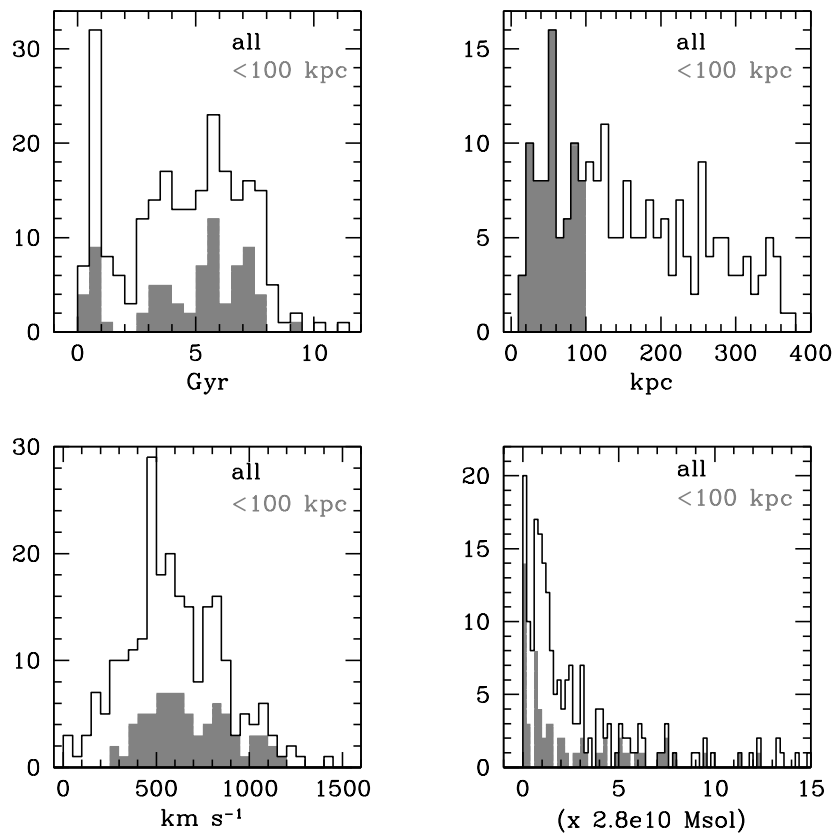}
\end{center}
\caption{General properties of all the encounters between a disc
galaxy and group members after it infalls with the most likely orbit. The
histograms combine information from the 12 REF.4g1tBA simulations, covering
different initial orbital parameters of group members. (top-left) Time of the
encounters since the infall of the disc galaxy onto the group, (top-right) radial
separation and (bottom-left) relative radial velocity between the disc galaxy and
each group member during encounters, and (bottom-right) total mass (DM+stars)
bound to each group member during encounters. Shaded areas show the properties of
the subset of encounters corresponding to a radial separation from the disc centre
of mass smaller than 100~kpc.}
\label{gral-props}
\end{figure*}

\begin{figure*}
\begin{center}
\includegraphics[width=65mm]{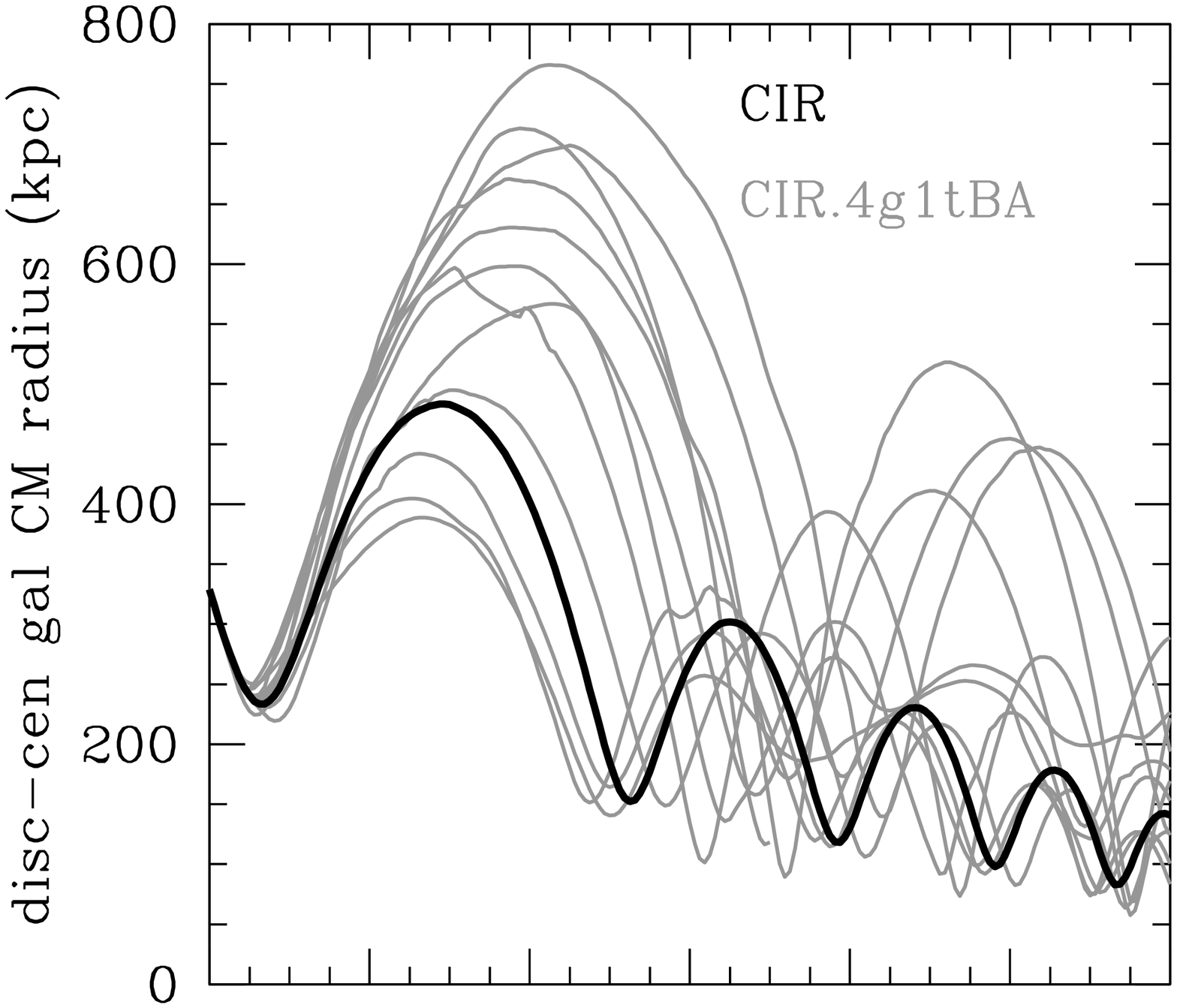}\hspace*{-11mm}
\includegraphics[width=65mm]{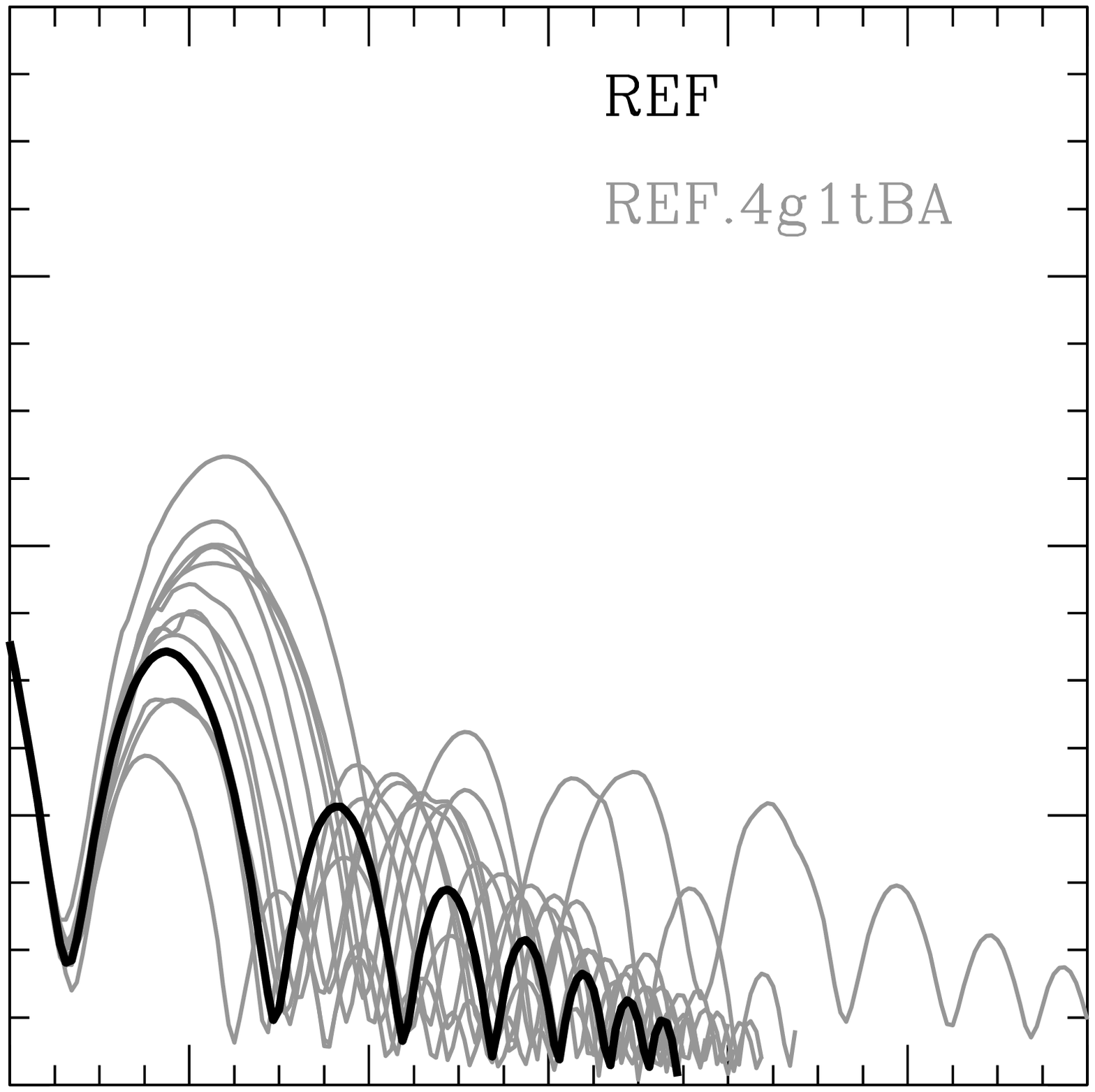}\hspace*{-11mm}
\includegraphics[width=65mm]{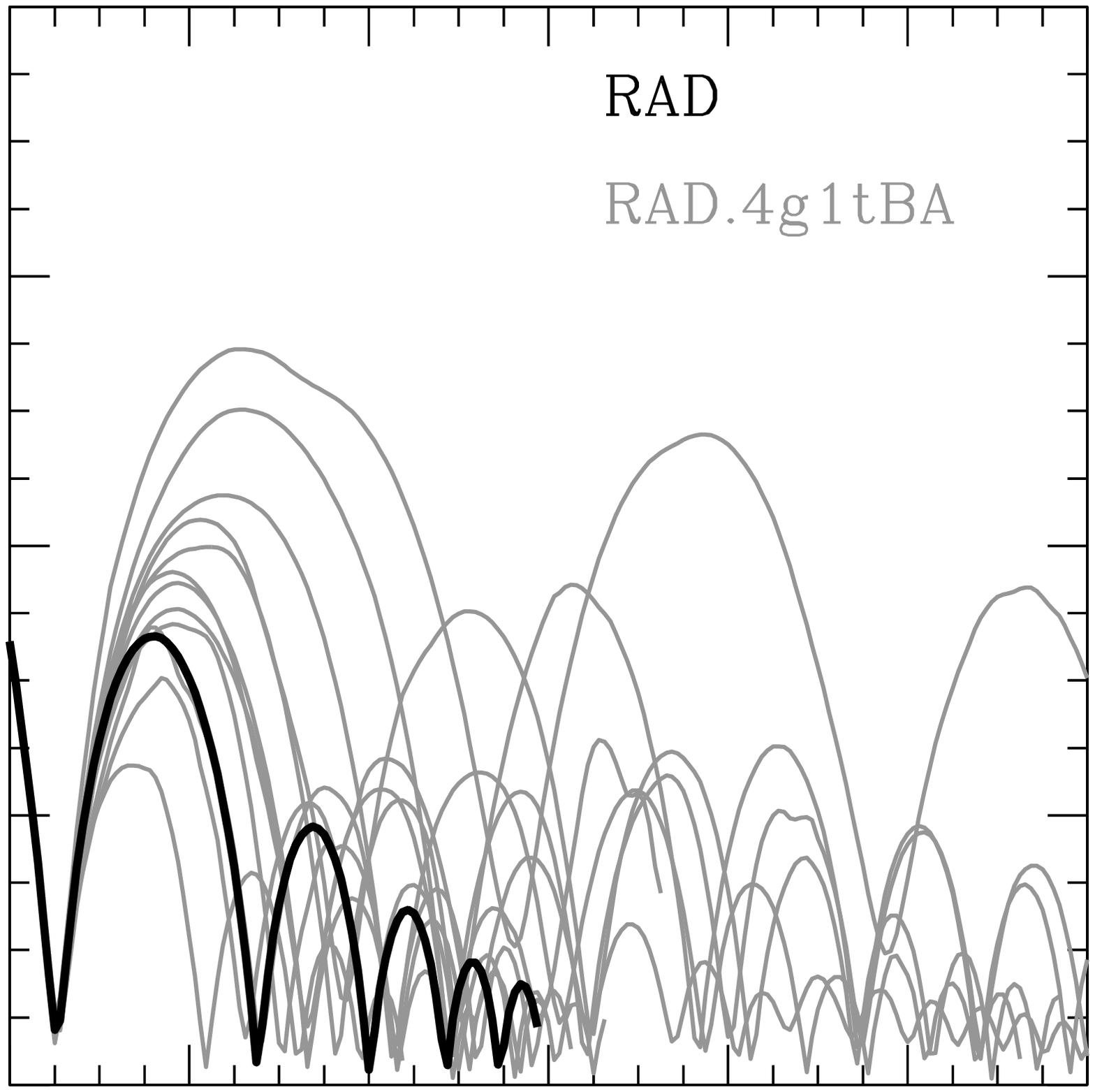}\vspace*{-11mm}\\
\includegraphics[width=65mm]{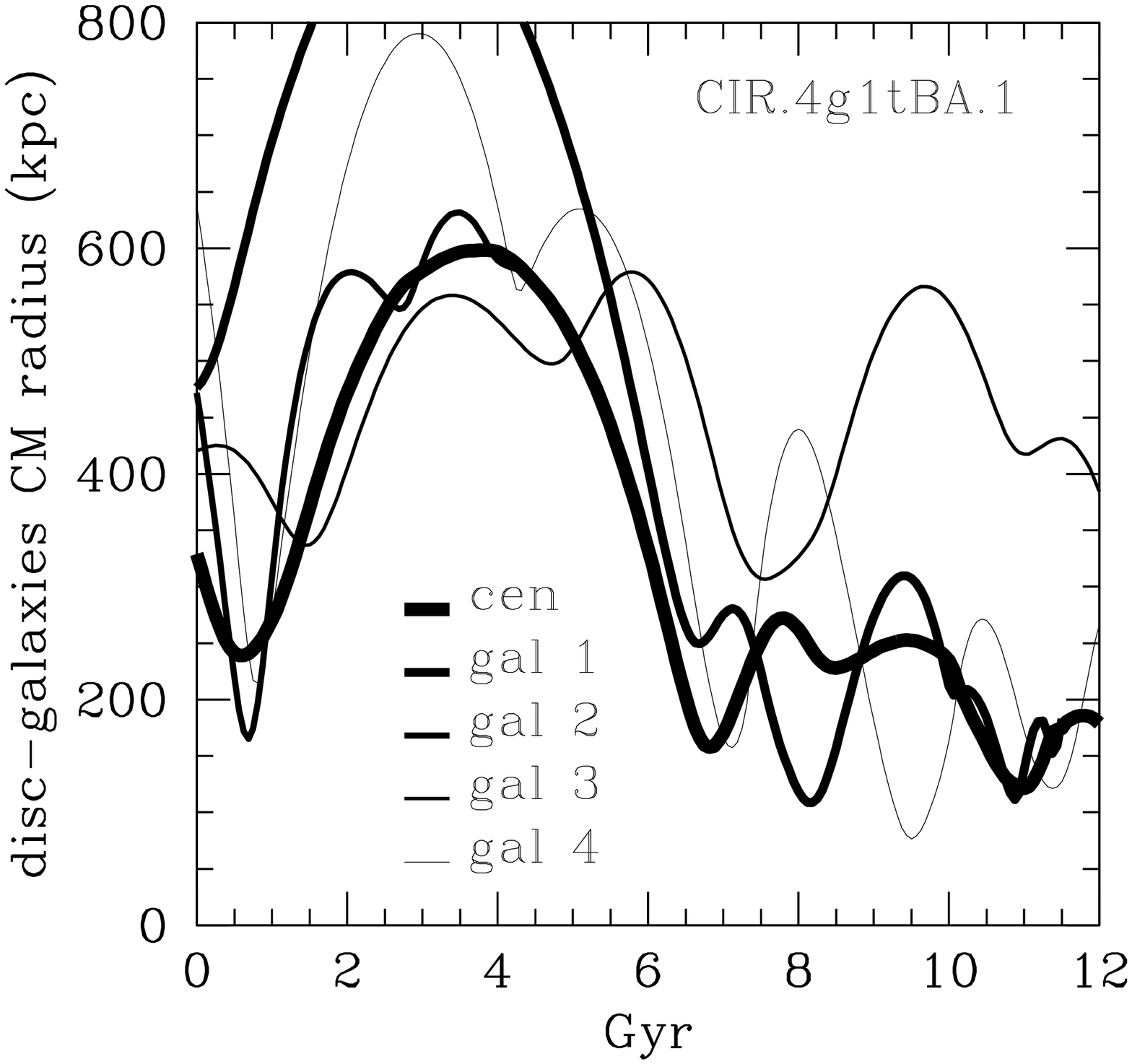}\hspace*{-11mm}
\includegraphics[width=65mm]{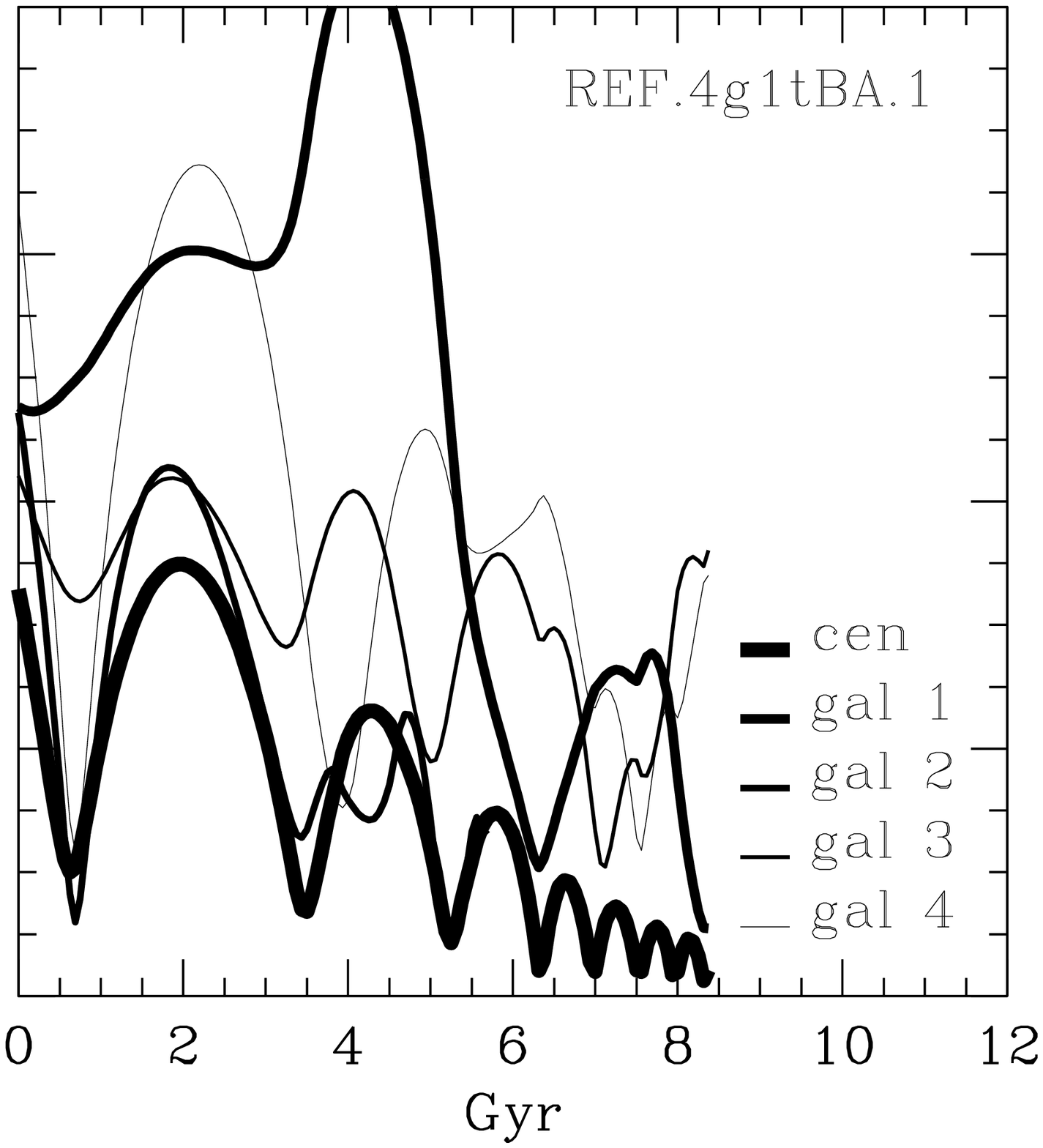}\hspace*{-11mm}
\includegraphics[width=65mm]{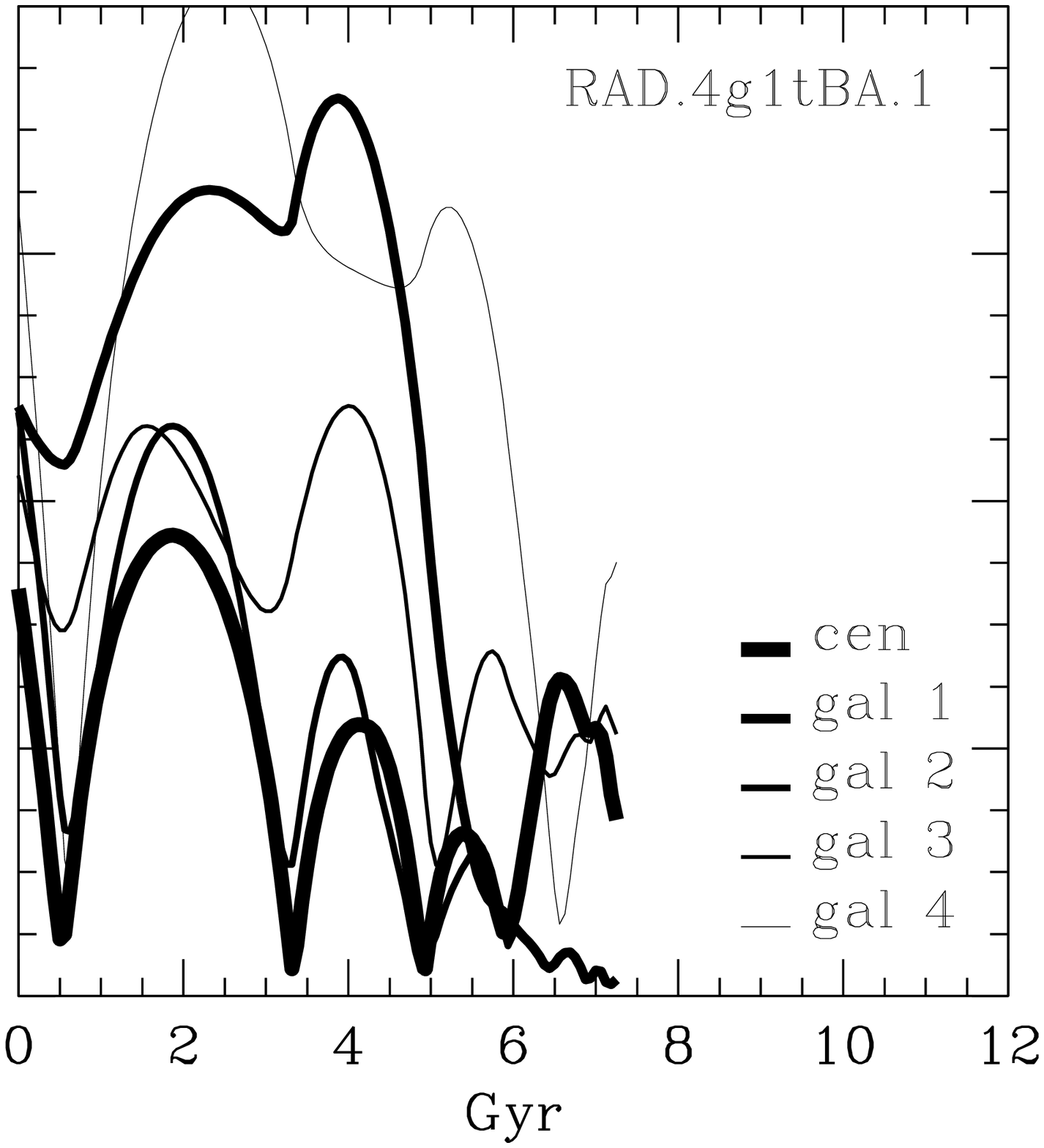}
\end{center}
\caption{ (top) Evolution of the radial separation between disc galaxies
infalling in the group and their respective central galaxy, for increasing
(initial) infalling orbital eccentricities: CIR, REF, RAD (see
Table~\ref{list-exper}). Each panel shows the evolution of the (12) different
initial orbital distributions for group members (grey), and the corresponding
case when no members are included in the group halo (black). (bottom) Evolution
of the radial separation between the disc galaxy and each galaxy in the group,
for increasing (initial) infalling eccentricities and the \emph{same} initial
orbital distribution for group members. All orbits are followed until either the
disc galaxy or the respective group member is disrupted.
}
\label{orb-massloss-disc}
\end{figure*}

\begin{figure*}
\begin{center}
\includegraphics[width=65mm]{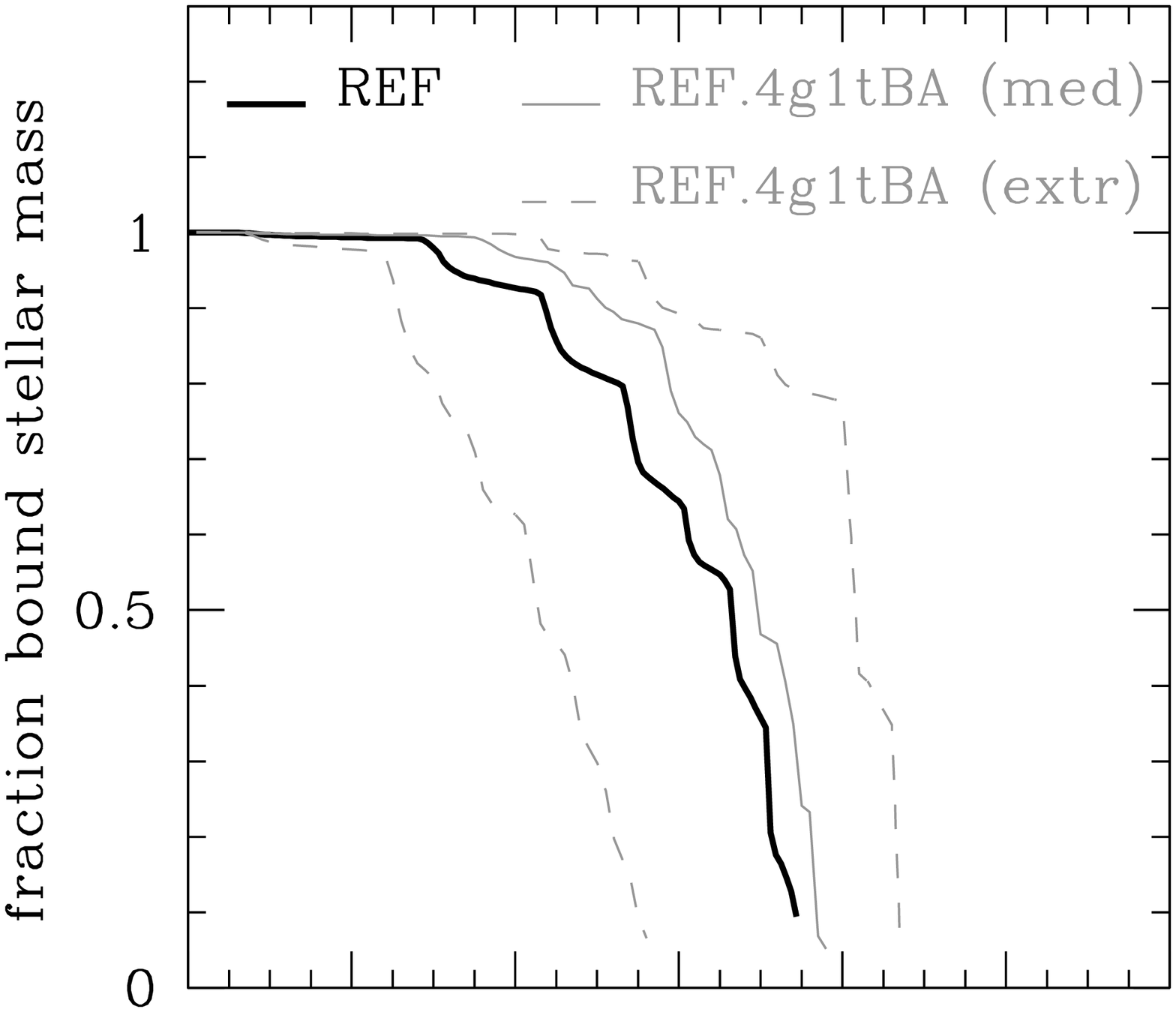}\hspace*{-11mm}
\includegraphics[width=65mm]{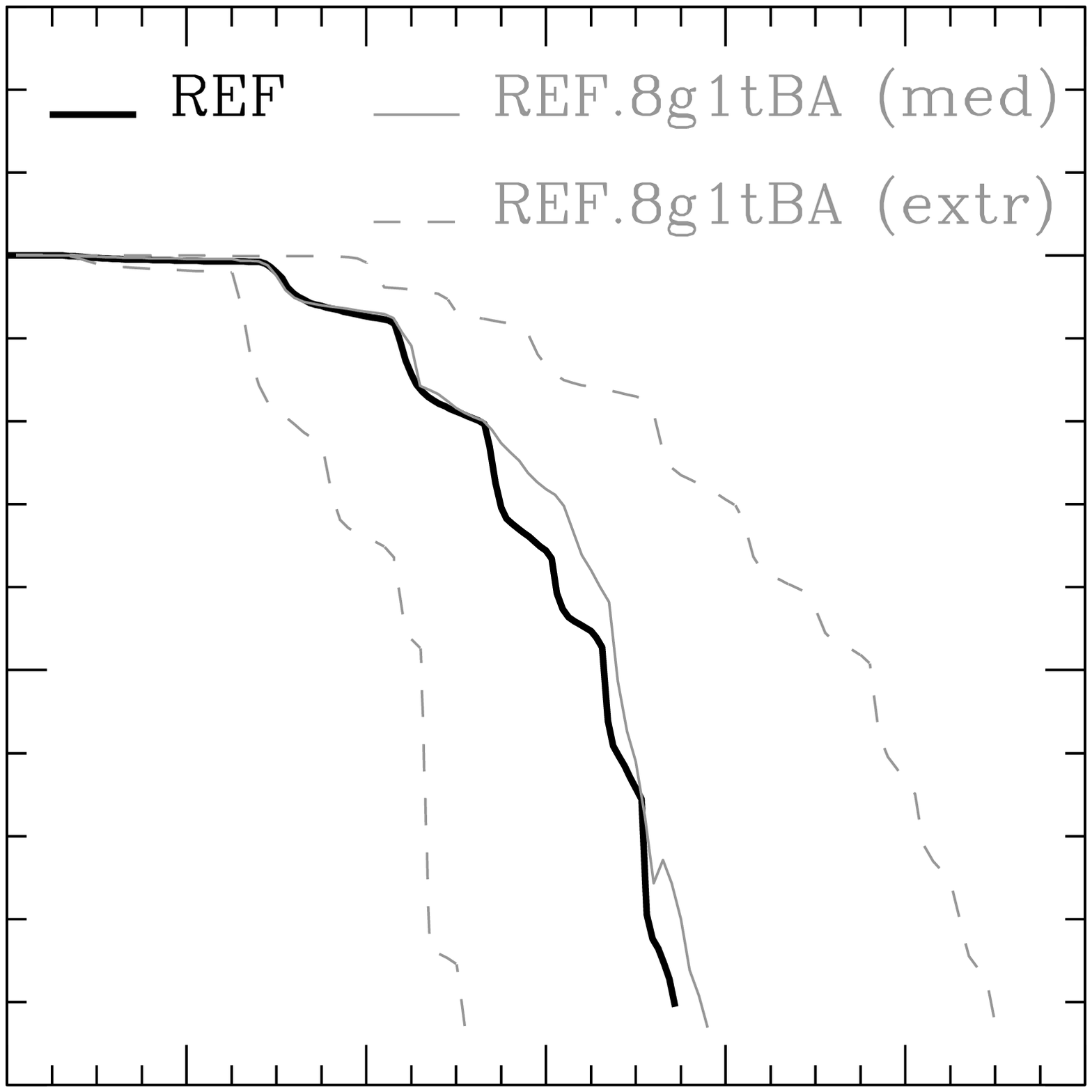}\hspace*{-11mm}
\includegraphics[width=65mm]{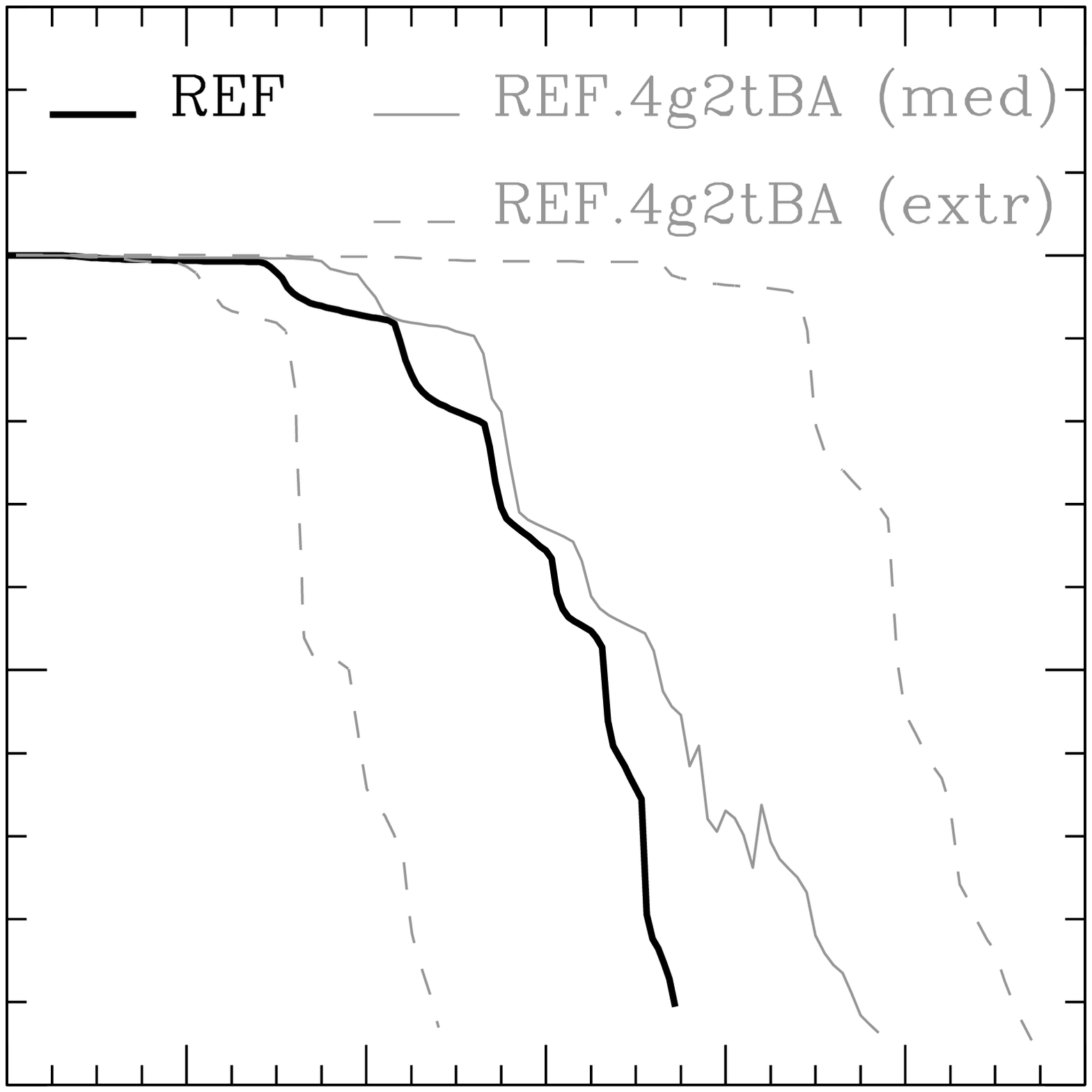}\vspace*{-11mm}\\
\includegraphics[width=65mm]{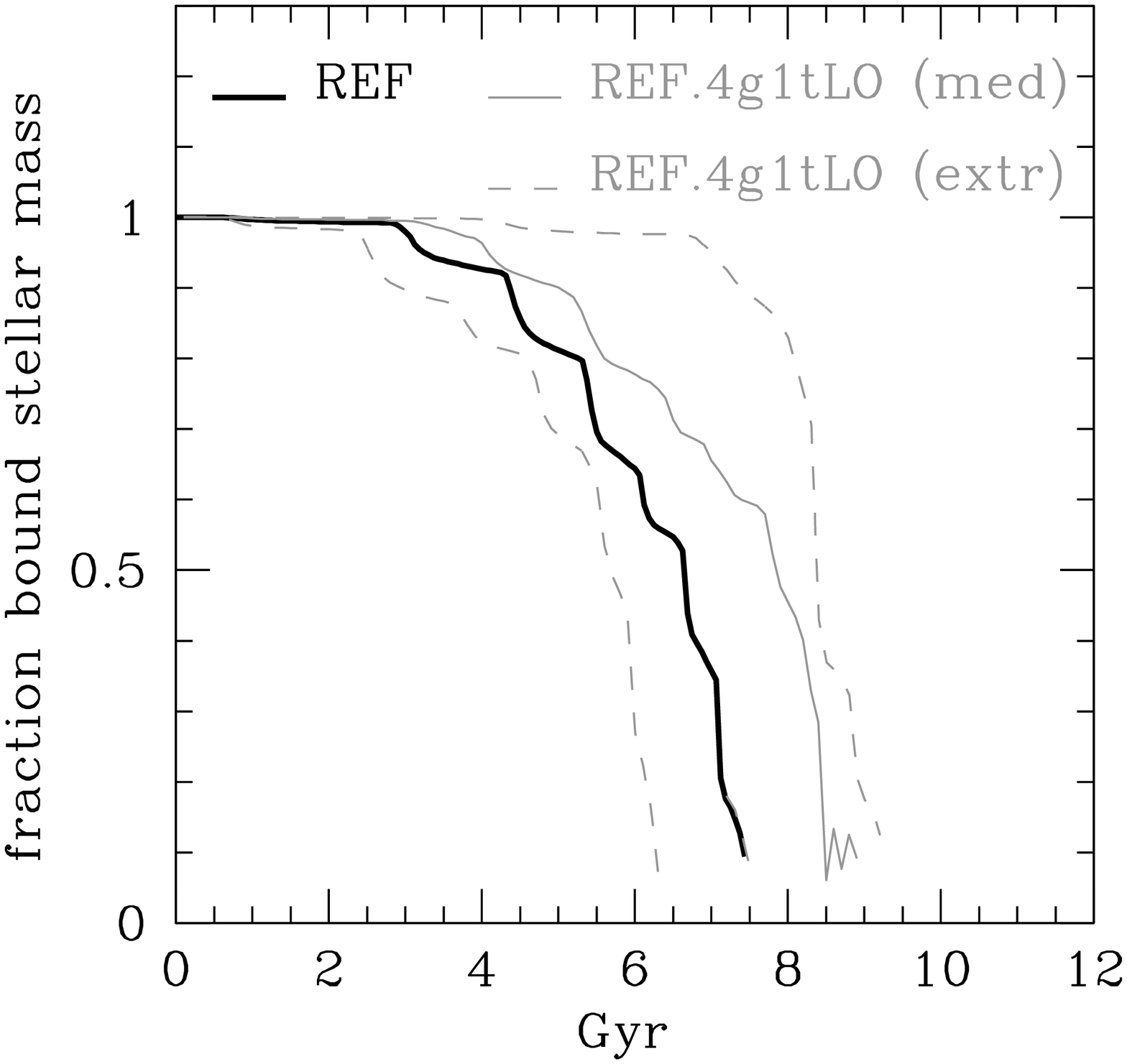}\hspace*{-11mm}
\includegraphics[width=65mm]{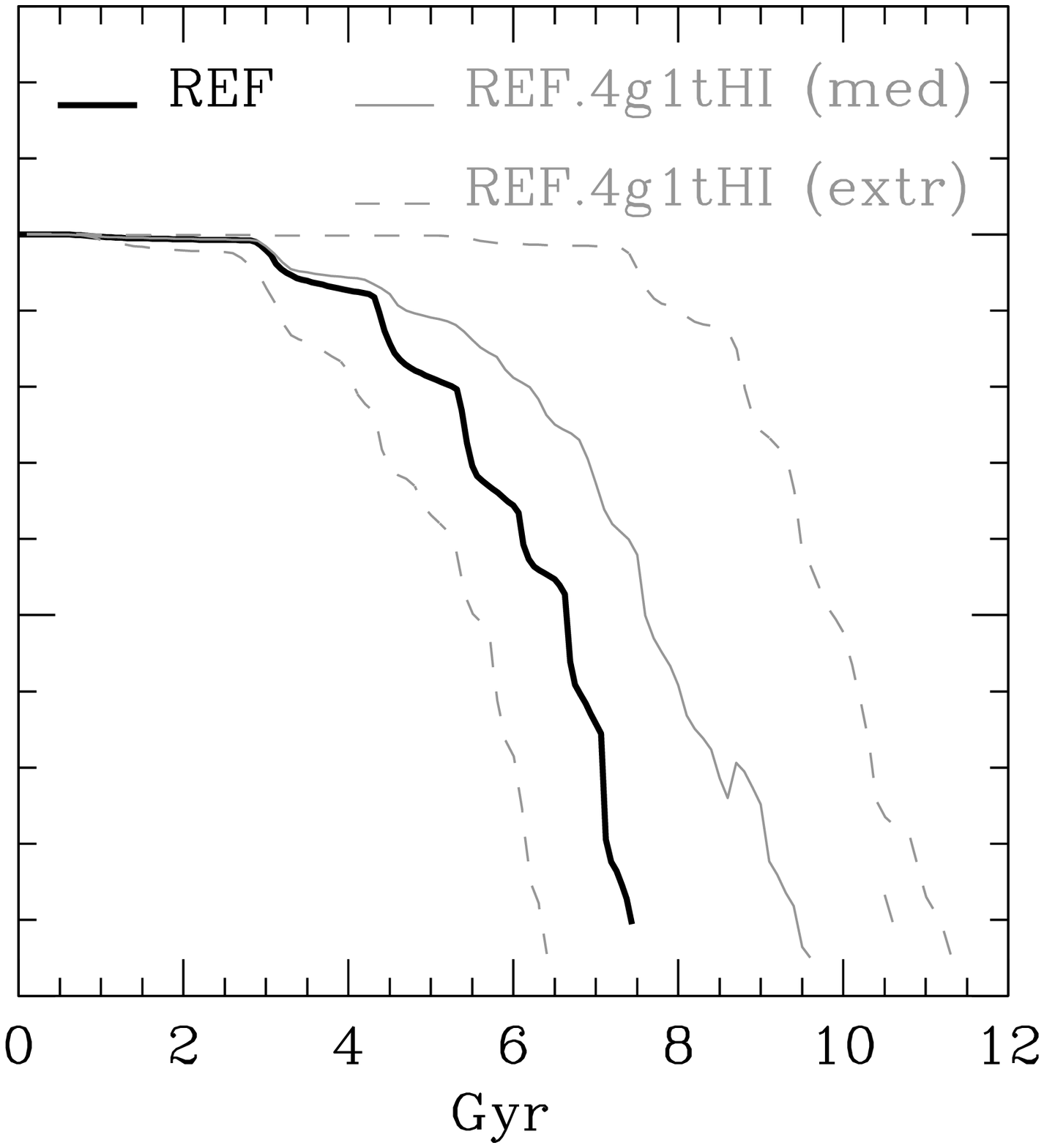}\hspace*{-11mm}
\includegraphics[width=65mm]{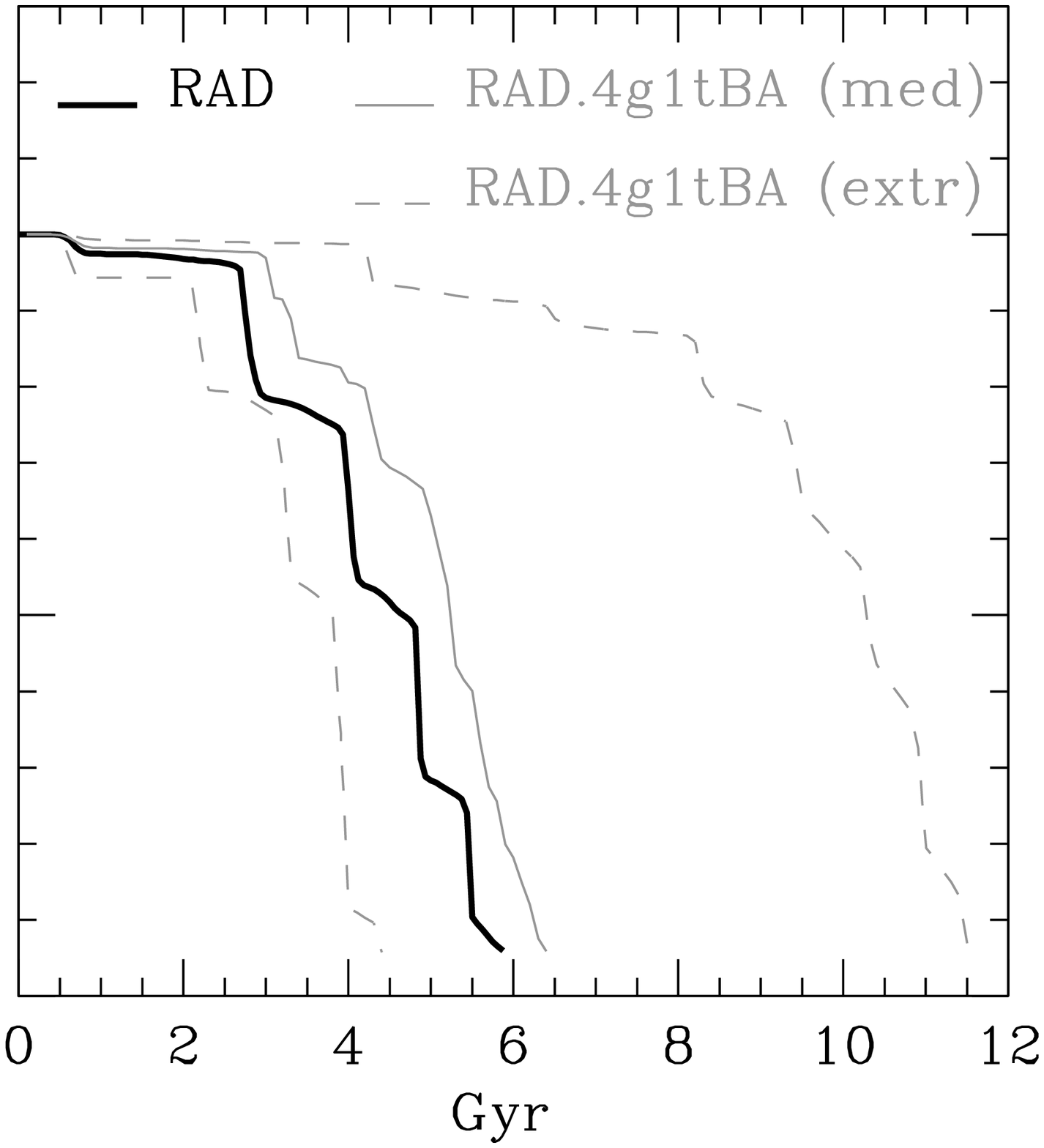}
\end{center}
\caption{Evolution of the fraction of stellar mass that remains bound to
disc galaxies (since their infall) in experiments covering different properties
of the group population (number of members, combined total mass, mass distribution),
and different (initial) infall eccentricities of disc galaxies.
In each case, different initial orbital distributions of group members induce
significant variations in the evolution of the fraction of stellar mass
that remains bound to discs. Each panel compares the median (grey solid) and
extreme cases (grey dashed) of the evolution due to the inclusion of group
members to the corresponding case when no members are present in the group halo
(black).}
\label{stellar-stripping}
\end{figure*}

\begin{figure*}
\begin{center}
\includegraphics[width=65mm]{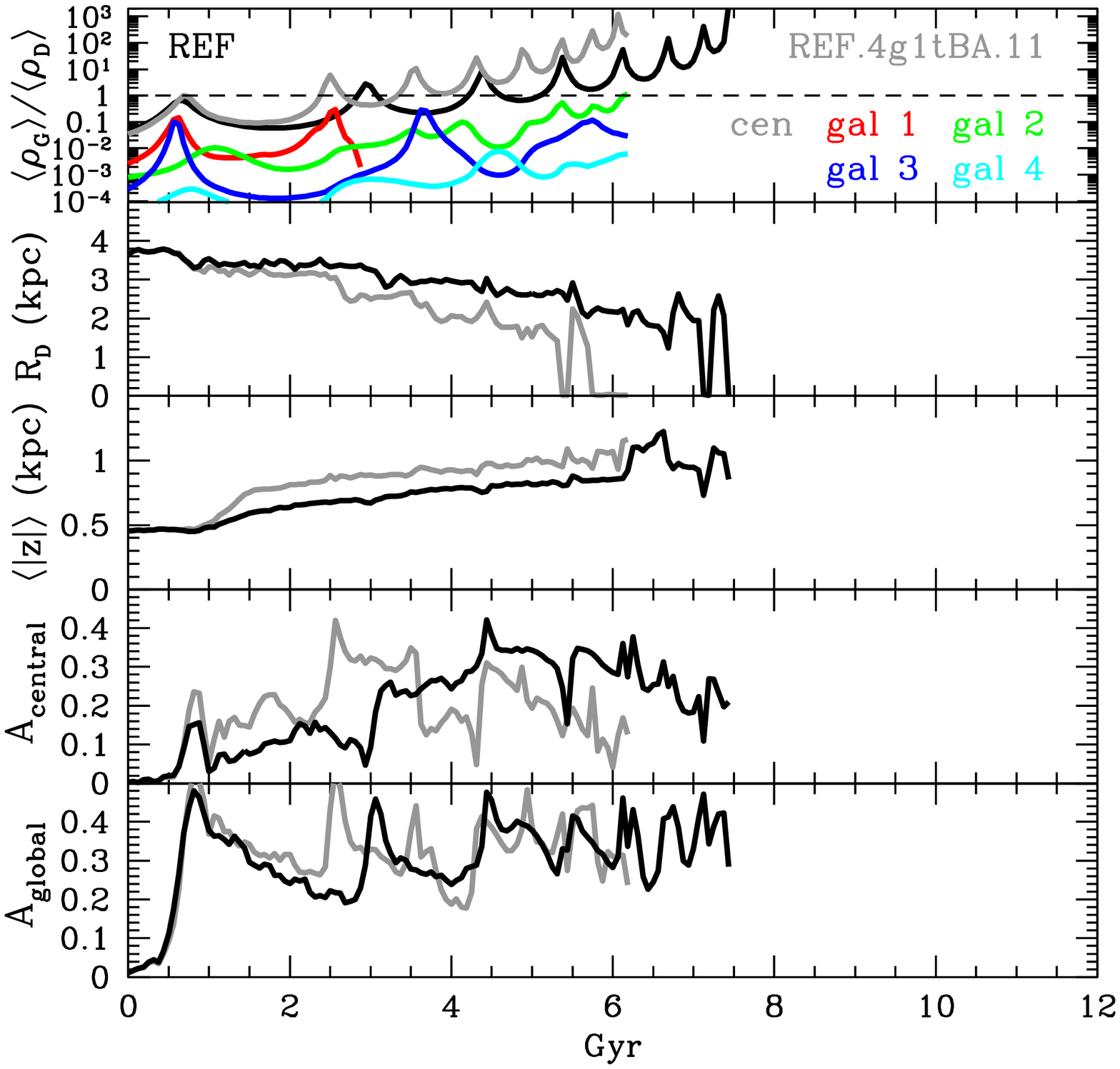}\hspace*{-11mm}
\includegraphics[width=65mm]{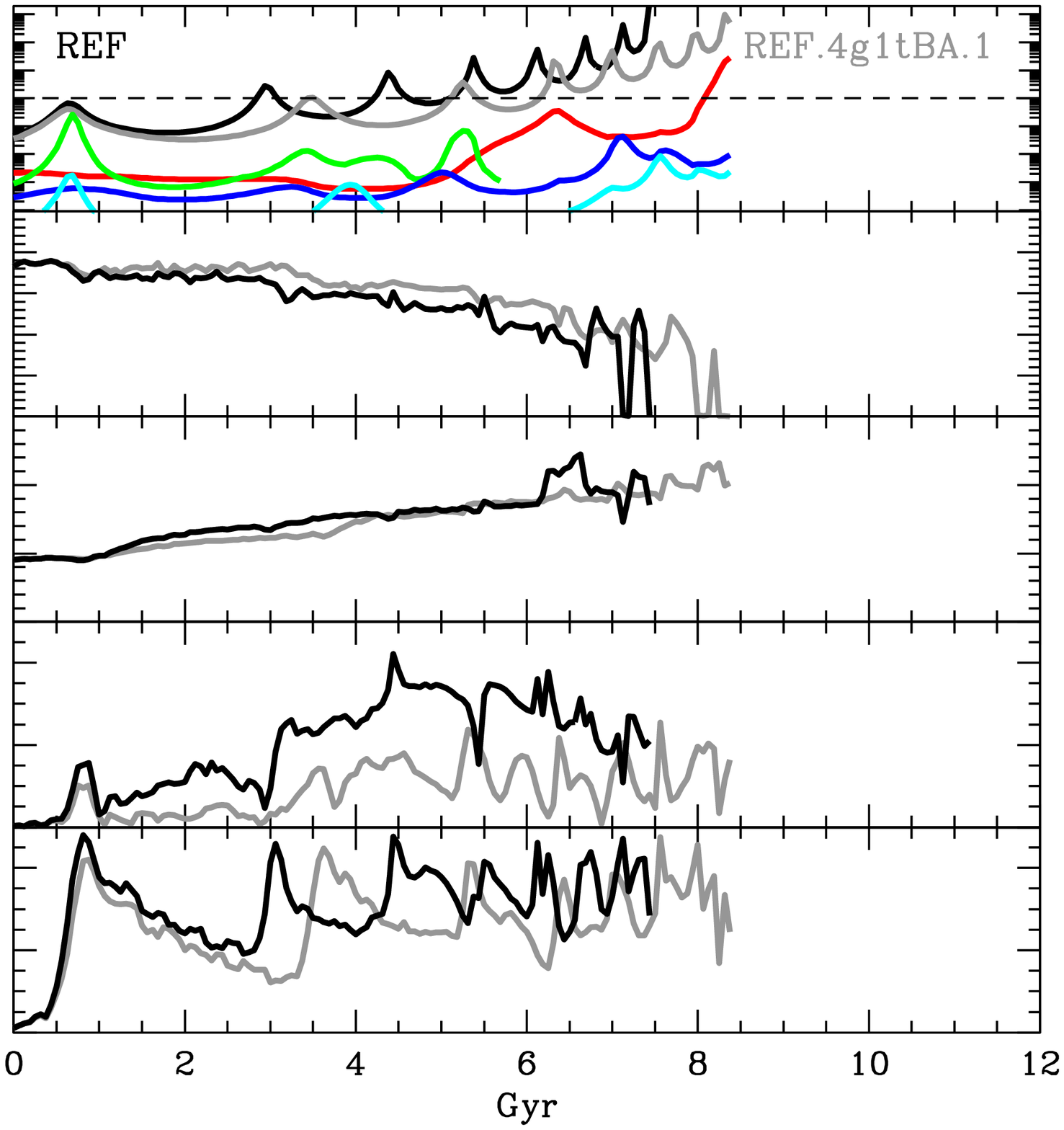}\hspace*{-11mm}
\includegraphics[width=65mm]{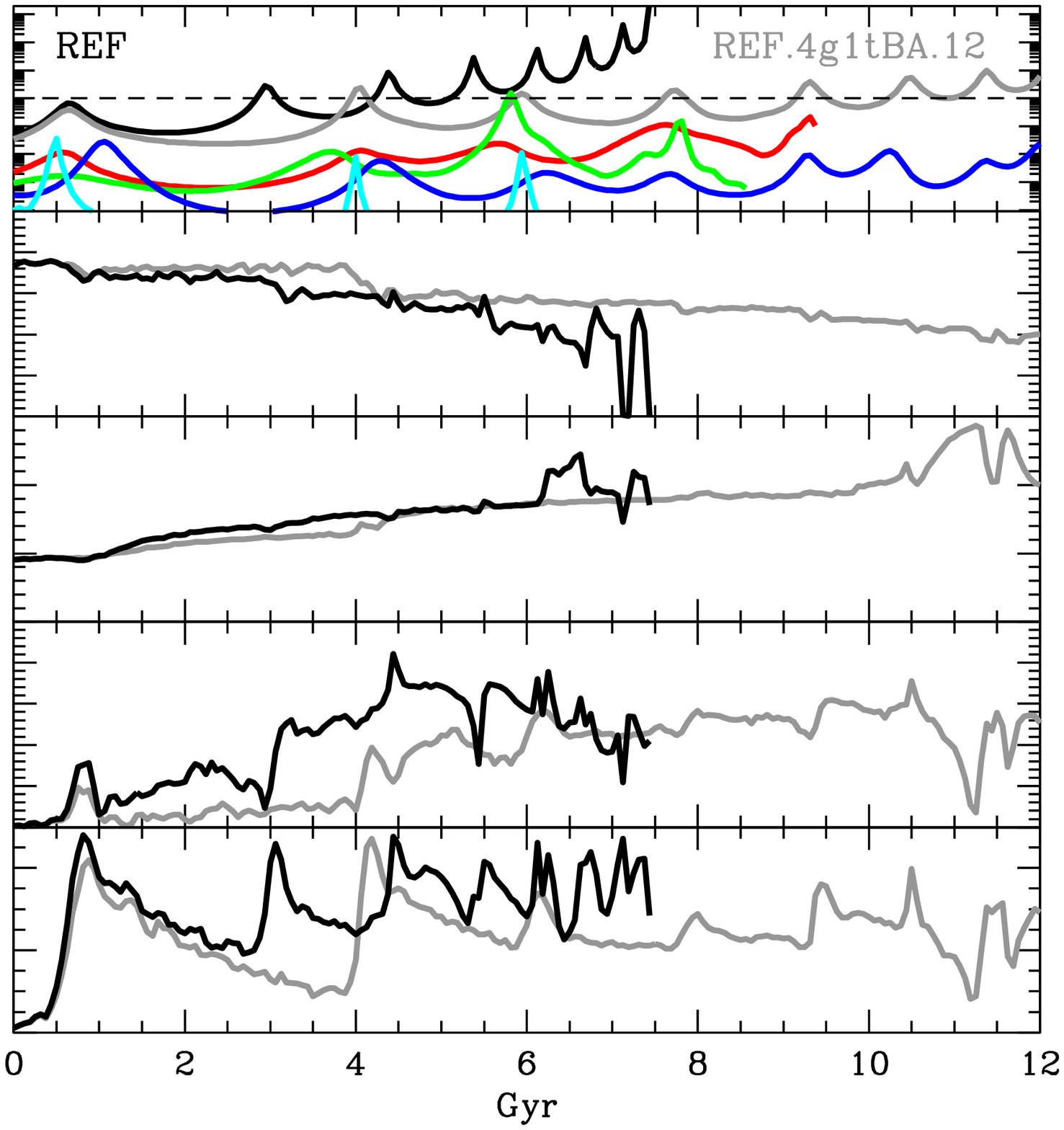} 
\end{center}
\caption{
Examples of the different structural evolution of a disc galaxy induced 
\emph{only} by different (initial) orbital distributions for group members, while 
keeping all other properties of the group population the same (number of members,
combined total mass, mass distribution). The structural evolution of the disc, 
infalling with the most likely orbital eccentricity, is presented in terms of its 
scalelength $R_{\rm D}$, mean thickness $\langle|z|\rangle$, and amplitude of 
both their central and global $m=2$ non-axisymmetries (grey). For comparison, the 
structural evolution of the corresponding disc when no members are present in the 
group halo is also shown (black). The evolution of the ``density contrast'' 
$\langle\rho_{\rm G}\rangle/\langle\rho_{\rm D}\rangle$ is also included to 
compare the tidal forces acting on the disc galaxy due to the central galaxy 
(grey), each group member (coloured, in decreasing order of total mass), and when 
no group members are present (black). The evolution of each property is followed 
until either the disc galaxy or any group member is disrupted.
}
\label{struc-evol-disc}
\end{figure*}

\section{Results}
\label{sec-descrip}

\subsection{General properties of galactic encounters}
\label{sec-gral-props}
In this paper, we define an encounter between the main disc galaxy and a group 
member as a \emph{local} minimum in the radial separation between their centres 
of mass, while both orbit within the group environment. Encounters are only 
registered while both the disc galaxy and a given group member retain more than 
5 percent of their initial stellar mass. In this way, a disc galaxy can 
experience several encounters with a single group member before either of their 
stellar components is disrupted.

Fig.~\ref{gral-props} shows an overview of the general properties of 235 
encounters that the disc galaxy experiences with group members, combining all 12 
simulations covering different initial spatial distributions of group members, 
for the experiment with the most likely disc galaxy's infalling orbit 
(REF.4g1tBA experiment).

We find that most encounters happen relatively late in our experiments, with a 
most likely time of occurrence of $t$$\sim$6~Gyr. This suggests that encounters 
take place mostly after both the main disc galaxy and group members have suffered
considerable mass stripping during their infalling orbits, which limits the 
damage caused by encounters on the disc galaxy. Note that a smaller fraction of 
encounters take place early in the simulations at $t$$\sim$1~Gyr. 
This is due to the random assignment of the initial velocities, which puts 
galaxies preferentially on radially inward orbits, increasing the probability of 
encounters in the inner regions of the group. 

We find that the range of radial separation between a disc galaxy and other group 
members during encounters is rather broad, reaching distances of $\sim$400~kpc. 
However, the most likely radial separation during encounters is found at 
$\sim$50~kpc. This is $\sim$~25 times the scale-length of disc 
galaxies at the most likely time of occurrence of encounters. 
``Very-close'' encounters ($<$20~kpc) and in particular mergers between galaxies are 
either rare or non-existent in our simulations. This is in contradiction with the 
common assumption that these events are frequent in the environment of groups 
(See Discussion).

The distribution of relative radial velocity between galaxies during encounters 
peaks around 500-600~km~s$^{-1}$. This shows that encounters are still relatively 
fast in comparison to the relative radial velocity expected in more massive 
clusters ($\sim$1000~km~s$^{-1}$). The most likely relative velocity of 
encounters in our simulations is consistent with estimations of the 
characteristic velocity dispersion of the group halo when approximated as an 
isothermal sphere. 

Finally, we find that the amount of \emph{total} (DM + stellar mass) mass bound to 
group members during encounters tends to be rather small, 
$\sim$2.8$\times$10$^{10}$M$_{\sun}$. In a typical encounter, a group member has 
a \emph{total} bound mass that is comparable to the mass bound to the stellar 
disc galaxy at the most common time for an encounter. We shall see that this, 
combined with the relatively large radial separation between galaxies during
encounters, contributes to weaker tidal forces on the disc from group members in 
comparison to the global tidal field.  

These general properties of encounters are similar to those of the rest
of our experiments, and they also apply when we restrict 
the analysis to ``close''-encounters only, i.e. those taking place within a 
radial separation $<$100~kpc (though of course the number statistics is reduced 
in this case).

A parallel can be drawn with \citet{knebe2006}, 
although the scope and methodology of their study are different from ours. They find 
that penetrating encounters between DM satellites in clusters formed in cosmological 
simulations are also rare and have high relative velocities that are comparable 
to the host one-dimensional velocity dispersion. Additionally, as we shall describe 
in Section~\ref{sec-stellar-stripping}, they also find that most of the mass loss 
experienced by satellites is due to interactions with the host potential as 
opposed to close-encounters.

\subsection{Orbital evolution of disc galaxies and separation to group members}
\label{sec-orb-evol}
We characterise the orbital evolution of galaxies by the evolution of the position
of the centre of mass of the stellar component of galaxies, considering only 
stellar particles that remain bound since the beginning of the simulations 
(see Section~\ref{sec-stellar-stripping}).

Fig.~\ref{orb-massloss-disc} (top) shows a comparison of the orbital evolution of 
the disc galaxy, for 12 different initial orbital distributions of group members 
and 3 initial orbital eccentricities of the disc. As a reference, the orbital 
evolution in the respective experiments without group members is also shown. The 
Figure shows that the inclusion of group members affects significantly the 
infalling orbit of the disc galaxy. In general, interactions with group members 
lead to an ``expansion'' of the infalling orbit of the disc, independently of its 
initial eccentricity. Interestingly, the orbital expansion is found to be the 
least significant for the case of the most likely eccentricity of infall. We 
argue that the expansion of the infalling orbits of disc galaxies does not come 
as a consequence of direct interactions between the disc galaxy and the other 
galaxies in the systems. Instead, this is likely due to the fact that by adding a 
galaxy population to the group halo-disc galaxy configuration, the total 
energy/angular momentum of the whole system is increased. In some cases, this 
causes a significant displacement of the densest region of the group halo while the 
disc galaxy is infalling toward it. Note that the global DM density profile of 
group haloes only shows a negligible evolution during our simulations.

Fig.~\ref{orb-massloss-disc} (bottom) shows the typical evolution of the radial 
separation between the disc galaxy and the central galaxy (thickest), and the group 
members (thinner), for different initial infall eccentricities 
of the disc galaxy, 
and for a fixed initial orbital distribution of group members. We find that 
closer encounters between the disc and group members take place when the orbit of 
the disc is more eccentric. This is to be expected since in such orbit the 
pericentres of a galaxy will be smaller than in less eccentric orbits, increasing 
the probability that the galaxy will pass close to the centre of the group, where 
group members are also infalling. 

\subsection{Stellar stripping of disc galaxies}
\label{sec-stellar-stripping}

We quantify the stellar stripping suffered by galaxies by computing the amount of 
bound stellar mass that galaxies retain during their evolution within the group 
halo. We apply the same algorithm as in \citet{villalobos2012}. Briefly, we:
(i) Consider all DM and stellar particles of the galaxy that were bound at the 
previous snapshot as bound at the current snapshot. If the current snapshot is 
the initial one, then all galaxy particles are considered bound by construction;
(ii) Compute the total bound mass of the galaxy and the velocity of its centre of 
mass;
(iii) Compute the binding energy of the particles that are considered bound using 
the updated velocity of the galaxy's centre of mass;
(iv) Retain only those particles that are still bound (i.e., with negative 
binding energy), and recompute the total mass of the galaxy;
(v) If the total bound mass from (ii) and (iv) has converged, then record it and 
go back to (i) for the next snapshot. If the total bound mass has not converged, 
then go back to (ii) using the bound particles found at step (iv).

Fig.~\ref{stellar-stripping} shows a comparison between the median and extremes 
of the stellar stripping evolution suffered by disc galaxies when group members 
in different orbital distributions are included in our simulations. As a 
reference, the figure also shows the disc stellar stripping for the corresponding 
experiments when group members are not included. Note that the figure only shows 
experiments where the disc galaxy is disrupted before 12 Gyr of evolution. This 
typically excludes 1 or 2 cases in each experiment which does not affect 
significantly the estimation of the ``median'' behaviour.  

We find that the inclusion of group members leads to a broad range of possible 
stellar stripping evolution in disc galaxies. In particular, the disruption time 
of identical stellar discs can vary within a range of $\pm$1 to $\pm$6~Gyr, with 
respect to the case when no group members are included. However, we find that the 
``median'' stellar stripping evolution (over different initial orbital 
distributions for group members) can be reasonably well described by the 
evolution of disc galaxies in simulations where group members are not included.
Even though the ``scatter'' in each experiment is relatively large, the lack of 
variation in the median evolution is encouraging for theoretical studies 
attempting to obtain prescriptions for stellar stripping evolution and merger 
timescales from controlled simulations of isolated mergers that do not account 
for the effect of other galaxies 
\citep[e.g.][]{boylan-kolchin2008,villalobos2013}. 

Interestingly, the ``median'' stellar stripping of stellar discs when group 
members are included in the experiments tends to be slower than when group 
members are not included. This is counter-intuitive as one would expect that close 
encounters accelerate the disruption of the infalling disc galaxy. The slower 
stellar stripping can be understood as a consequence of the ``expansion'' of the 
infalling orbit of disc galaxies, as explained in Section~\ref{sec-orb-evol}. 
This delays the orbital passages of disc galaxies around the dense central region 
of the group, where they are exposed to stronger tidal forces. This also 
highlights the fact that in our simulations stellar stripping in disc galaxies is 
mostly driven by interactions with the global environment (which continuously grow 
in mass by the disruption of group members) rather than by direct interactions 
between galaxies in the group (these are weak, as discussed in 
Section~\ref{sec-gral-props}). 

Note that these results are also valid for the disruption of the DM halo of disc 
galaxies and are similar between experiments covering different orbital 
eccentricities of disc galaxies, number of group members (within a factor of 2), 
total mass enclosed by group members (within a factor of 2), and different stellar 
mass distributions for group members (being most of them either less or more 
massive than the stellar discs). The experiments show that the amount of 
``scatter'' in the mass stripping mostly depends on the combined total mass of 
group members. This is expected since a more massive population of group members 
would inject a larger amount of angular momentum to the system\footnote{Assuming a 
fixed group total mass, this corresponds to cases where the global DM content has 
a small contribution to the group total mass \citep[see][]{aceves2013}. Note that 
in these cases the variability in the properties of infalling disc galaxies is 
particularly high given the larger amount of angular momentum being transferred by 
more massive group members.}. Note that different morphologies of infalling 
satellites could represent an additional source of ``scatter'' in their mass 
stripping evolution \citep[see][]{chang2013}. Interestingly, the case when a disc 
galaxy infalls in a more radial orbit is associated with an asymmetric scatter in 
its stellar stripping. 

\subsection{Structural evolution of stellar discs}
\label{sec-struct}
As in \citet{villalobos2012}, we study the structural changes of disc galaxies as 
they orbit within a group environment. We start by first centring the discs on 
their centre of mass and aligning them in such a way that the Z-axis is defined 
by their rotation axes. Then, structural properties have been computed in 
concentric rings, 1~kpc wide (out to 20~kpc from the disc centre), considering 
only stars that remain bound to the disc galaxy at a given time, and that are 
located within 3~kpc from the midplane. We refer the reader to 
\citet{villalobos2012} for a detailed description of the procedure.

Fig.~\ref{struc-evol-disc} illustrates the changes in the structural evolution 
of a disc galaxy that are introduced \emph{only} by different (initial) orbits of 
group members (while keeping the same number of members, their combined total 
mass and mass distribution). The disc galaxy infalls with the most likely orbital 
eccentricity. The structural evolution of the stellar disc is shown in terms of 
its scale-length $R_{\rm D}$, mean thickness $\langle|z|\rangle$, and amplitude 
of both central and global $m=2$ non-axisymmetries $A_{\rm central}$ (bar), 
$A_{\rm global}$ (tidal arms). For comparison, the evolution of the stellar disc 
in the corresponding experiment without group members is included. The 
Figure also shows the evolution of the ``density contrast'' 
$\langle\rho_{\rm G}\rangle/\langle\rho_{\rm D}\rangle$, to compare the 
tidal forces acting on the disc galaxy due to the central galaxy and each group 
member. This quantity corresponds to the ratio between the mean mass density of 
the central galaxy (or a group member) within the distance to the disc galaxy, 
and the mean mass density of the disc galaxy within 10 initial scalelengths. Both 
mean densities only include DM and stars that remain bound at a given time. The 
evolution of the ``density contrast'' due to the central galaxy when no group 
members are present is also included.

We find that the inclusion of a population of group members does not have 
necessarily a destructive effect on the structure of the infalling disc galaxy on 
top of the evolution induced by the global tidal field. In fact, it is not 
uncommon to find that the inclusion of group members actually slows down the 
structural evolution of discs induced by the global tidal field. This can be seen 
not only in terms of the disc's scalelength and  mean thickness, but interestingly 
also in the reduced formation of central bars\footnote{In the context of our 
simulations, the observed environmental dependence between bars/bulges likelihood 
with environment \citep[e.g.][]{skibba2012} would appear to be linked to the 
presence of gaseous components in group galaxies.}. The Figure also shows that 
tidal forces on a disc galaxy by group members are in most cases weaker than those 
induced by the central galaxy. This is consistent with the description of 
Fig.~\ref{gral-props}, where encounters were found to be mostly distant in 
comparison to the extension of discs and to take place mostly after group members 
lose a significant amount of their initial mass content. This indicates that the 
global tidal field of a group plays a much more relevant role in the evolution of 
galaxies in comparison to that of close-encounters. 

In Fig.~\ref{struc-evol-disc} the structural evolution of discs and the 
estimation of tidal forces are followed only until either the stellar component 
of the disc galaxy or that of a group member has been disrupted. We find that 
identical disc galaxies (infalling with the same orbit) can survive longer, 
when the more massive group members also survive for a longer 
time. This points toward an important transfer of angular momentum from group 
members (especially the most massive ones) to the whole system as they are 
affected by dynamical friction within the group. In fact, our results suggest 
that the overall influence of group members on an infalling disc galaxy comes 
mainly indirectly by modifying the global tidal field of the group as they are 
disrupted, and by adding angular momentum to the whole system as they suffer 
dynamical friction.

Figs.~\ref{scal-evol-disc} and \ref{thick-evol-disc} in the Appendix show 
the evolution of the scalelength and mean thickness of disc galaxies in all 
our experiments. We find that the median evolution can be well described by
simulations that do not include group members, while the inclusion of group members 
introduces a significant scatter.

\section{Discussion}
\label{sec-discussion}
In the previous Section we have shown that, within the parameter space explored 
with our simulations and the assumptions we have adopted, close-encounters 
between galaxies in groups are rare. Additionally, according to our simulations, 
close-encounters in groups have a limited \emph{direct} effect on the galaxies 
involved. Instead, most of the effect of group members on an infalling disc 
galaxy comes \emph{indirectly} via modifications to the global tidal field of the 
group, as group members are tidally stripped and/or merge. Note that this implies 
a potentially high variability in the way a galaxy can be affected by other group 
members, even when only the orbital parameters of group members are 
different. 

\subsection{On correlations between galaxy properties and host halo mass}

Previous studies have shown a correlation between several observables of satellite 
galaxies (age, metallicity, quiescent fraction), and the mass of their host haloes  
\citep[e.g.][]{pasquali2010,delucia2012,wetzel2012}. These correlations are found 
to be relatively weak for massive satellites, while properties of less massive 
galaxies show a clearer dependence on host halo mass. As shown in 
Section~\ref{sec-descrip}, even within host haloes of the same mass a particular 
galaxy can experience very different evolutionary paths, being strongly dependent 
on the evolution of other group members and how they affect their common tidal field.
The high variability in the evolution of group galaxies found in our simulations 
would likely weaken any underlying correlation with halo properties. Even though our 
measurements cannot be easily translated into the observed correlations, it is 
striking to consider that the scatter found in our simulations could be a 
conservative estimation. Both rapid variations in the global potential through halo 
mass growth and clustered accretion of group members could further increase the 
scatter (see Section~\ref{subsec-caveats}). 

Alternatively, \citet{hou2013} find no significant correlation between the quiescent 
fraction of galaxies with the dynamical mass of groups within the 
10$^{13}$--10$^{14.5}$ M$_{\sun}$ mass range, in catalogues from the Sloan 
Digital Sky Survey (SDSS), the Group Environment and Evolution Collaboration (GEEC) 
and the high redshift GEEC2 sample out to $z$$\sim$1\footnote{As stated by
\citeauthor{hou2013}, this apparent contradiction with previous studies is caused by 
different cuts in halo mass adopted to explore underlying correlations.}. 
Interestingly, they also find that the quiescent fraction is \emph{lower} in groups 
\emph{with} substructure for low mass galaxies ($\log[M_{*}/M_{\sun}] < 10.5$). 

At face value, both results are consistent with the outcomes of our simulations, 
\emph{if} it is assumed that stronger gravitational interactions between a galaxy 
and/or the global field are correlated to higher fraction of quiescent galaxies.
The unclear correlation with halo mass found by \citeauthor{hou2013} is inline with 
the large ``scatter'' that group galaxies should exhibit in their properties 
according to our simulations. The lower fraction of quiescent galaxies observed by 
\citeauthor{hou2013} in groups with substructure can be understood in terms of the 
connection between the evolution of a given galaxy and that of other group members, 
as shown in Fig.~\ref{struc-evol-disc}. The Figure shows that the longer time the 
more massive group members/substructures survive within the group (i.e. are 
detectable), the less effective the global tidal field acting on an infalling galaxy 
is, causing it to retain its (stellar and eventually gaseous) mass content/structure 
for a longer time. However, it is important to highlight that, as opposed to our 
group members, in \citeauthor{hou2013} substructure lies well beyond the viral 
radius of the halo, which makes difficult a direct comparison of their effects. 

\subsection{Comparison to the effect of the global tidal field from previous simulations}

\subsubsection{Start of structural transformations in galaxies within groups}
In \citet{villalobos2012} we study the effect of only the global tidal field of 
the group environment on the evolution of galaxies, by means of controlled 
$N$-body simulations similar to those presented in this work. In our 
previous study, we found that accreted disc galaxies start suffering a 
significant structural transformation due to the global tidal field only after 
the mean density of the group, within the orbit of the galaxy, is $\sim$0.3--1 
times the central mean density of the galaxies. After including the effect of 
other galaxies in the group environment, we find that this result remains a good 
approximation, as illustrated for a few cases in Fig.~\ref{struc-evol-disc}.
This correspondence highlights the usefulness of relatively simple and 
less costly simulations to characterise the ``mean'' evolution of disc galaxies 
being accreted onto a group. Specifically, these results can be used to assist 
the implementation of particular physical processes in analytic and semi-analytic 
models, e.g. as done in recent studies on the formation of the intra-cluster light in 
hierarchical galaxy formation models \citep{contini2014}, and on  
environmental influence on quenching star formation (Hirschmann et al. 2014, 
submitted).

\subsubsection{On the formation of S0 galaxies in groups}
In \citet{villalobos2012} we also find that the global tidal field of a group 
alone is inefficient at either inducing the formation of central bulges in 
pre-existing stellar discs (i.e. in place before the accretion onto the group) or 
enhancing pre-existing bulges. These results can have important 
implications for the formation of S0 galaxies in group environments. S0 galaxies 
are characterised by having little gas content, practically no signs of spiral 
arms, and often a prominent central bulge. Their formation processes is currently 
unknown, although they are usually considered to be the end product of spiral 
galaxies affected by environmental processes 
\citep{dressler1980,solanes1992,vandokkum1998}. Increasing observational evidence 
points toward the group environment as the characteristic environment for the 
formation of these galaxies \citep{wilman2009,just2010}. In 
\citeauthor{villalobos2012} we concluded that, if S0 galaxies are preferentially 
formed in group environments, then their prominent bulges could not be produced 
by the effect of group tidal forces alone. This implies that bulges of S0 
galaxies formed in groups would be composed mostly by young stars. In this study, 
we show that also the combined effect of global tidal field and close-encounters 
is not efficient at inducing/enhancing bulges of old stars in stellar discs. This 
is illustrated for a few cases in Fig.~\ref{struc-evol-disc} in terms of the 
amplitude of central instabilities (also confirmed by examining the evolution of 
the surface density profiles of each galaxy in our experiments). It is 
interesting to note that the effect of other group members can actually 
inhibit/delay the formation of central instabilities in comparison to the effect 
of the global tidal field (Section~\ref{sec-struct}). The implications of our 
simulations are consistent with recent observational evidence based on 
absorption-line index gradients \citep{bedregal2011} and spectroscopic bulge-disc 
decomposition \citep{johnston2012} for a sample of S0 galaxies in the Fornax 
cluster, indicating that in those galaxies bulges are younger than the 
stellar discs\footnote{However, most of well spectro-photometrically studied S0 
galaxies are high mass ones, which might have a different formation path
with respect to low mass satellite S0 galaxies, as shown by \citet{wilman2012,wilman2013}.}. 

\subsubsection{On prescriptions of merger timescales for galaxies within groups}

By means of isolated controlled simulations of mergers similar to those presented 
in this study, \citet{villalobos2013} introduce a modification to the 
prescription of merger timescales obtained by \citet{boylan-kolchin2008} from
comparable simulations. This modification consists in the inclusion of an 
explicit dependence on the redshift at which a galaxy is accreted onto a host 
halo and it is motivated by the evolution of halo concentration with redshift. 
The modified prescription offers an improvement up to $\sim$20 $t_{\rm dyn}$ in 
the prediction of merger timescales up to $z=2$, in absence of interactions with 
other substructures.

In Fig.~\ref{stellar-stripping} we show that, when the effect of close-encounters 
with other group members are included in the simulations, the ``median'' stellar 
mass loss of a galaxy (over experiments with different initial orbital parameters 
for otherwise the same set of group members) can be reasonably well approximated 
by the mass loss experienced by the same galaxy only under the influence of the 
global tidal field of the group. This shows that when close-encounters are 
included in the simulations, the ``median'' merger time of a galaxy can be well 
predicted by a relatively simple prescription based only on the effect of the 
global tidal field (and dynamical friction). The inclusion of other group members 
introduces a dispersion, which appears to be mostly a function of the total mass 
of the galaxy population (or the corresponding global DM contribution to the group 
total mass), their mass distribution, and number of members. Such a dispersion 
could be introduced in galaxy formation models to explore its effect. Presumably, 
this would increase the scatter of predicted properties for group galaxies.

\subsection{Caveats}
\label{subsec-caveats}
Regarding the initial conditions adopted for our simulations, disc galaxies 
resemble both in mass and radial extension a slightly less massive Milky Way-like 
galaxy at $z$=0. Instead the radial extension (and concentration) of the DM halo 
of the group environment for its quoted virial mass resembles a halo at $z$=1. 
This configuration is set by design as we have attempted to maximise the effect 
of close-encounter between galaxies by placing them in a relatively smaller 
volume. Note however that the structure of the simulated group is still within 
ranges that are consistent with cosmological simulations. We have also carried 
out test simulations after scaling the properties of \emph{both} disc galaxies 
and the group DM halo at several redshifts, obtaining an even smaller effect on 
the evolution of galaxies due to close-encounters.
        
Our simulations do not account for the hierarchical mass growth that the group DM 
halo might experience after the disc galaxy under study has been accreted. In 
this way we do not consider possible rapid variations in the global potential due 
to the accretion of massive substructure. In addition, by assigning random 
positions and velocities to group members within a halo, our simulations could 
over-smooth the distribution of galaxies, thereby reducing the effect of close 
encounters. We estimate that the effect of hierarchical mass growth of the group 
halo on the properties of the disc galaxy would be similar to the effect 
introduced by the inclusion of group members in the simulations. That is, accreted 
substructure would modify the global tidal field of the group by either adding 
mass to its centre via mergers and/or by adding energy/angular momentum to the 
system. Depending on the mass of the accreted substructure, this would translate 
into an additional ``scatter'' in the evolution of the properties of galaxies 
around the ``median'' evolution expected within a group halo that does not 
contain other galaxies and does not experience mass growth.  

Finally, our simulations do not consider cases where galaxies are accreted as part 
of ``sub-groups'' onto a group halo, as it would be expected in the context of 
hierarchical evolution of galaxy haloes. In this way, we neglect a possibly 
significant effect of close-encounters at low velocity dispersions of ``sub-group'' 
scales \citep[see][]{feldmann2011}.

\section{Conclusions}
\label{sec-conclusions}
In this work, we study the evolution of disc galaxies inhabiting a group environment, 
using controlled collisionless $N$-body simulations of isolated mergers. Our goal is 
to estimate the relative contributions of both the global tidal field and 
close-encounters between group members to the evolution of disc galaxies. We probe a 
parameter space that covers a number of relevant aspects of the galaxy-group 
interaction. Regarding the disc galaxies, we explore different initial inclinations 
(with respect to the disc rotation), different orbital eccentricities (consistent 
with cosmological simulations), and the presence of a central stellar bulge in the 
disc. Regarding the population of satellite galaxies in the group, we explore 
different number of members (consistent with observations of groups), variations in 
their stellar mass content (also consistent with observations), and variations in the 
total mass of the satellite population. Our fiducial disc galaxy resembles a 
bulge-less, slightly less massive Milky Way.

Our main results are the following:
\begin{itemize}
\item
Close-encounters in a group environment with the most likely number of members
are found to be rare and gravitationally weak. Most of the encounters between 
galaxies occur at separations $\sim$50~kpc and relatively late in the 
simulations, when group members have already lost a sizeable fraction of their 
initial mass, reducing significantly their capacity to affect gravitationally the 
disc galaxy. Thus, in our simulations, the \emph{direct} effect of close-encounters 
between galaxies in groups is much less relevant than the influence of the global 
tidal field of the group.

\item
Within the group environment, the influence of other members on a given galaxy is 
found to be mostly \emph{indirect} by altering the global tidal field, as group 
members transfer to it mass (via tidal stripping) and energy/angular momentum (via 
dynamical friction). The mass added to the central region of the group shortens the 
merger timescale of an infalling galaxy, causing it to experience more structural
transformations and mass stripping. On the other hand, the addition of energy/angular 
momentum provokes a displacement of the densest region of the group, causing a delay 
in the evolution of an infalling galaxy, as the time between pericentric passages 
about the group centre increases.

\item
We find that the evolution of general properties of disc galaxies (such as merger 
timescale, stellar mass loss, scale-length, mean thickness) caused by other group 
members is highly variable, depending mostly on the initial orbital parameters 
(positions and velocities) of group members. However, the ``median'' evolution of 
disc galaxies is found to be reasonably well approximated by the effect only due 
to the global tidal field (i.e. in absence of close-encounters). Interactions with 
group members introduce a ``scatter'' around the ``median'' evolution of disc 
galaxies, whose amplitude depends mostly on the combined total mass of the group 
members, their number and mass distribution.

\item
Counter-intuitively, disc galaxies can sometimes be \emph{less} affected by 
environmental effects after interacting with group members, i.e. galaxies can 
retain their initial structure and mass content for a longer time. The high 
variability of the properties of disc galaxies due to the influence of group 
members could also wash out underlying correlations with environmental properties, 
such as the group mass. 

\item 
The effect of the global tidal field combined with close-encounters between galaxies 
is inefficient at inducing or enhancing the formation of central bulges in stellar 
discs present at accretion time. This confirms our previous conclusion that, if S0 
are formed from spiral galaxies preferentially in group environments, then their 
often large central bulges should contain mostly relatively young stars in comparison 
to their stellar discs. This result is found to be consistent with recent observations 
of S0 galaxies in the Fornax cluster.    

\item 
Finally, we find that prescriptions based on simulations that do not account for 
close encounters (e.g. those done to obtain merger timescales) remain valid and can 
indeed be used in the framework of galaxy formation models. More sophisticated 
implementations of those prescriptions could include a scatter dependent e.g. on the 
total stellar mass, mass distribution and number of nearby galaxies.
\end{itemize}

In the first two papers of this series we have studied the influence of the global 
tidal field and that of close-encounters on the evolution of disc galaxies within 
group environments, estimating their relative contributions. Thus far, we have 
focused on their effect on the stellar content of galaxies that is already present 
when disc galaxies are accreted onto a group. In the third paper of this series, 
we will examine the contribution of the gaseous components of disc galaxies in groups.

\section*{Acknowledgements}

We thank D. Wilman for useful suggestions on the manuscript. \'AV and GDL acknowledge 
financial support from the European Research Council under the European Community's 
Seventh Framework Programme (FP7/2007-2013)/ERC grant agreement n. 202781. This work 
has been partially supported by the PRIN-INAF 2009 Grant ``Towards an Italian Network 
for Computational Cosmology'', by the European Commission's Framework Programme 7, 
through the Marie Curie Initial Training Network CosmoComp (PITN-GA-2009-238356), 
and by the PD51 INFN Grant. We acknowledge the CINECA award under the ISCRA 
initiative, for the availability of high performance computing resources and support.

\bibliographystyle{mn.bst}
\bibliography{mn-jour,biblio}

\appendix

\section{Scalelength and thickness evolution of stellar discs}
\label{app-struct}
\begin{figure*}
\begin{center}
\includegraphics[width=65mm]{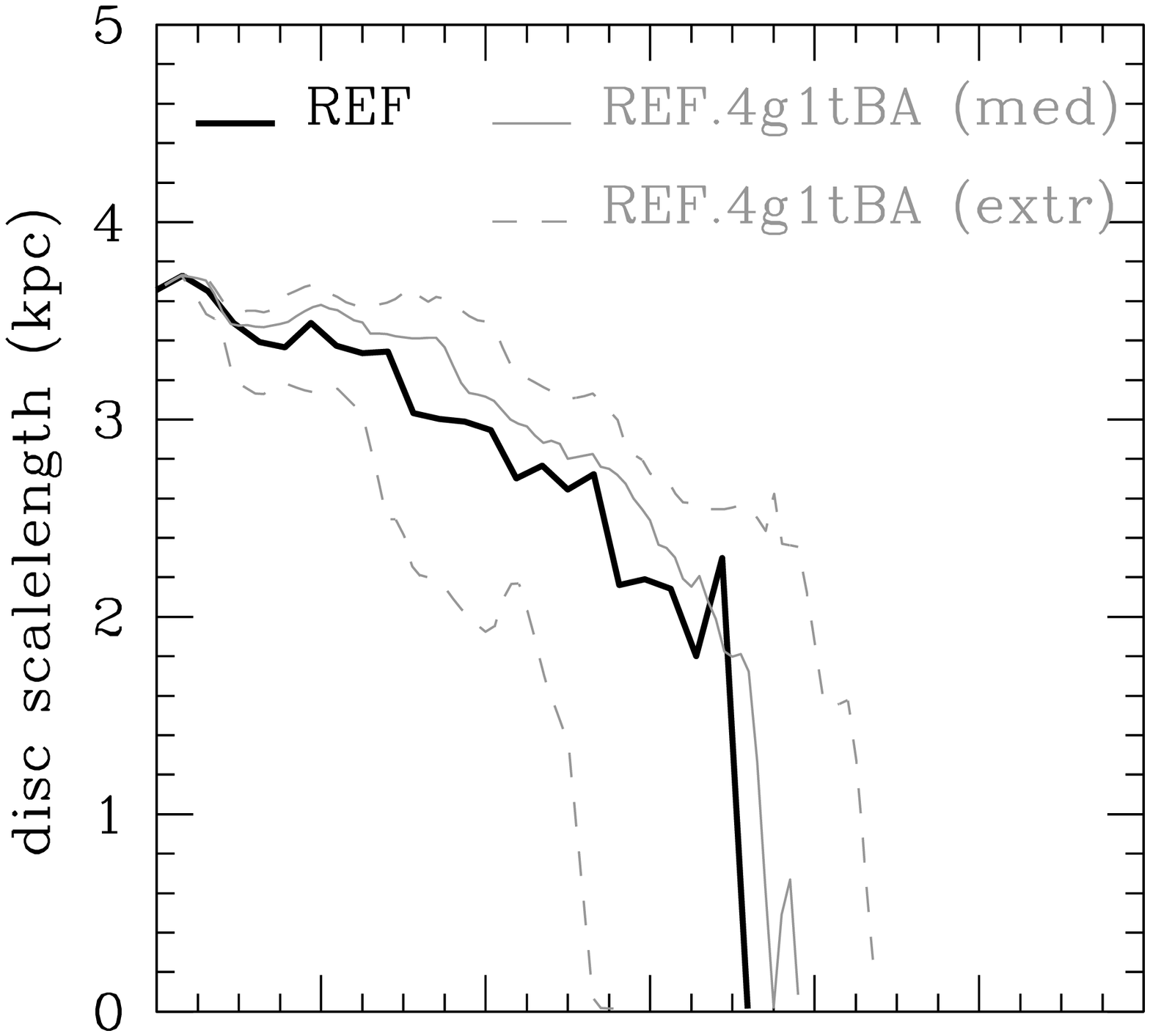}\hspace*{-11mm}
\includegraphics[width=65mm]{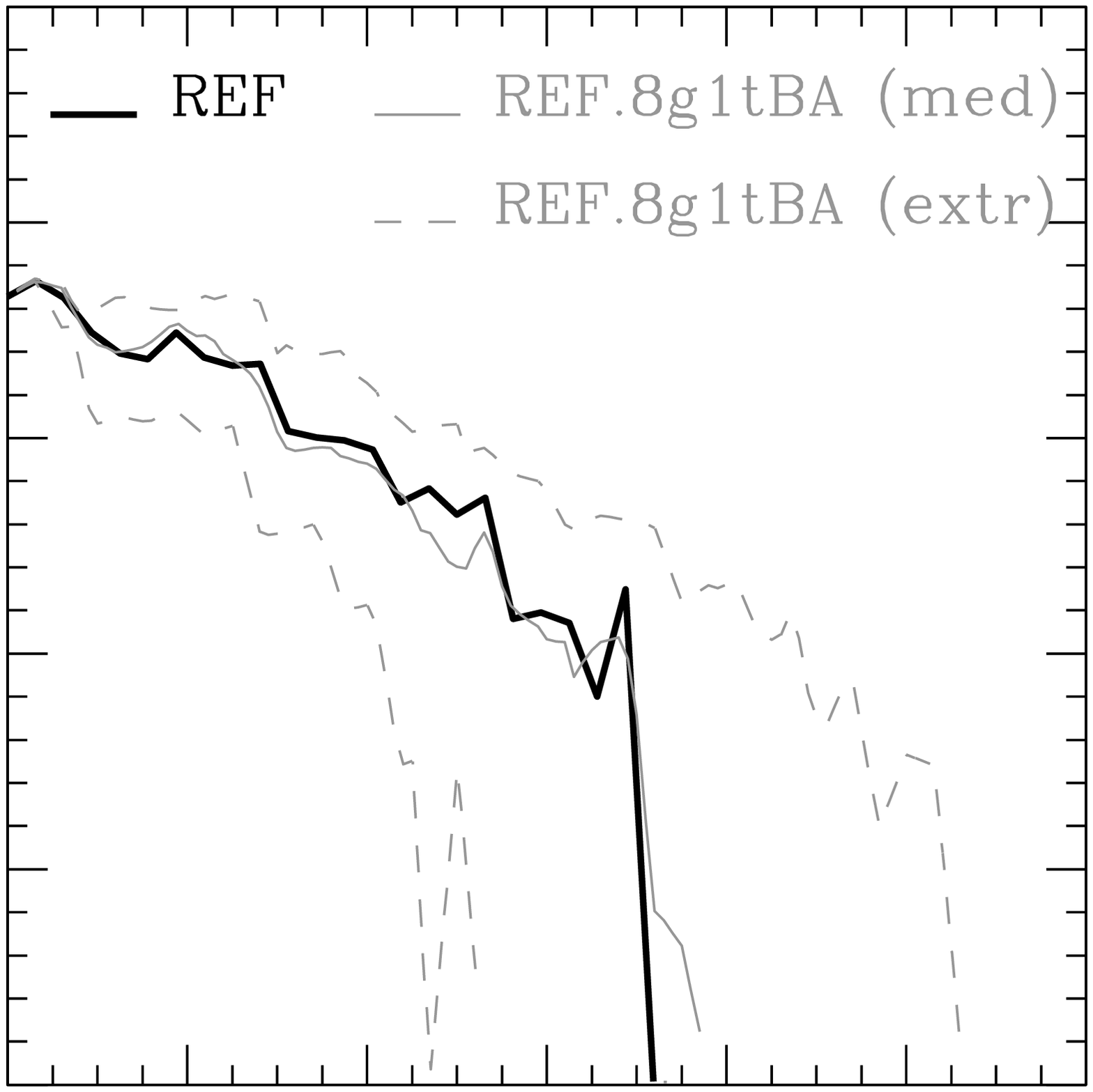}\hspace*{-11mm}
\includegraphics[width=65mm]{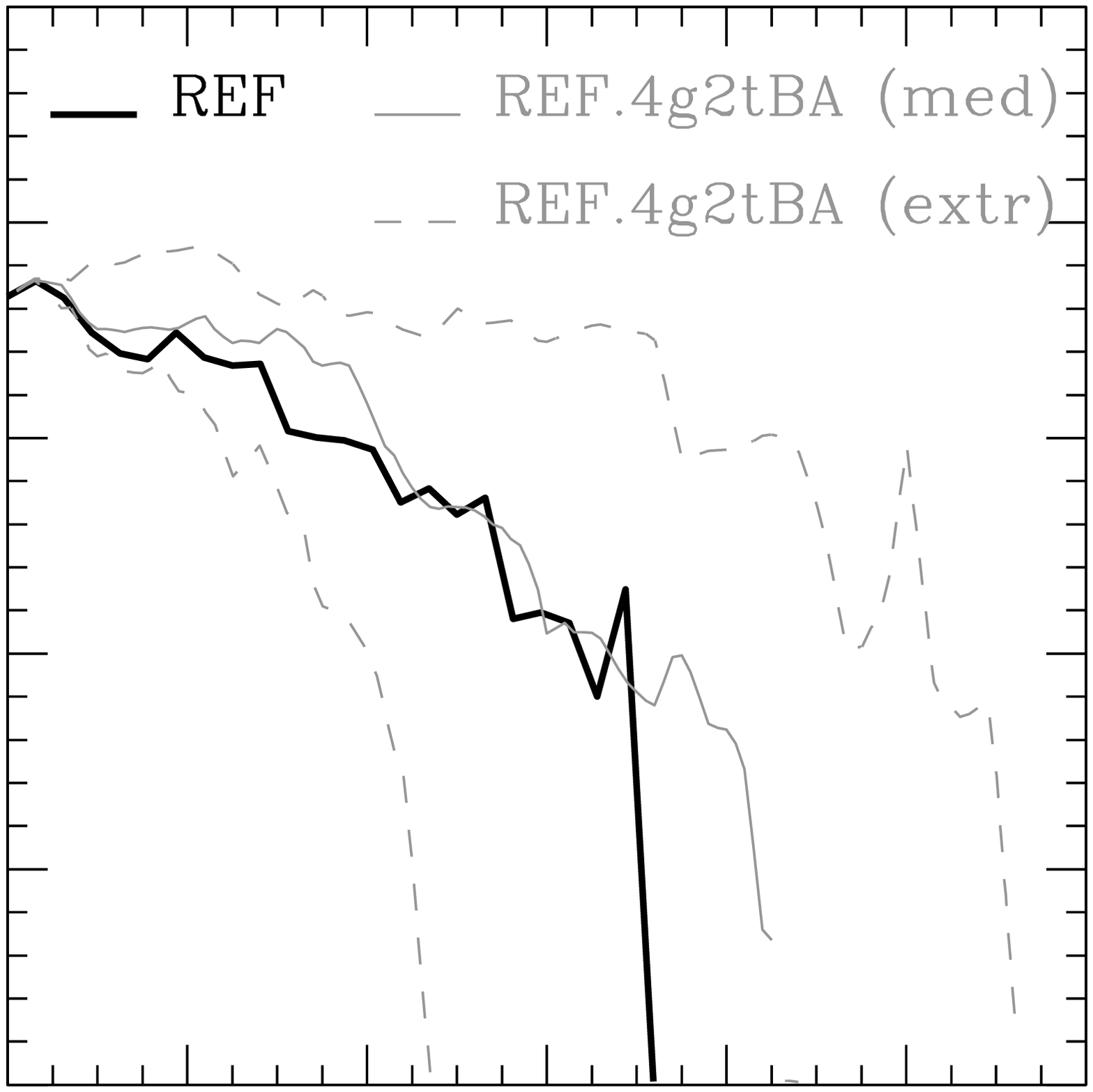}\vspace*{-11mm}\\
\includegraphics[width=65mm]{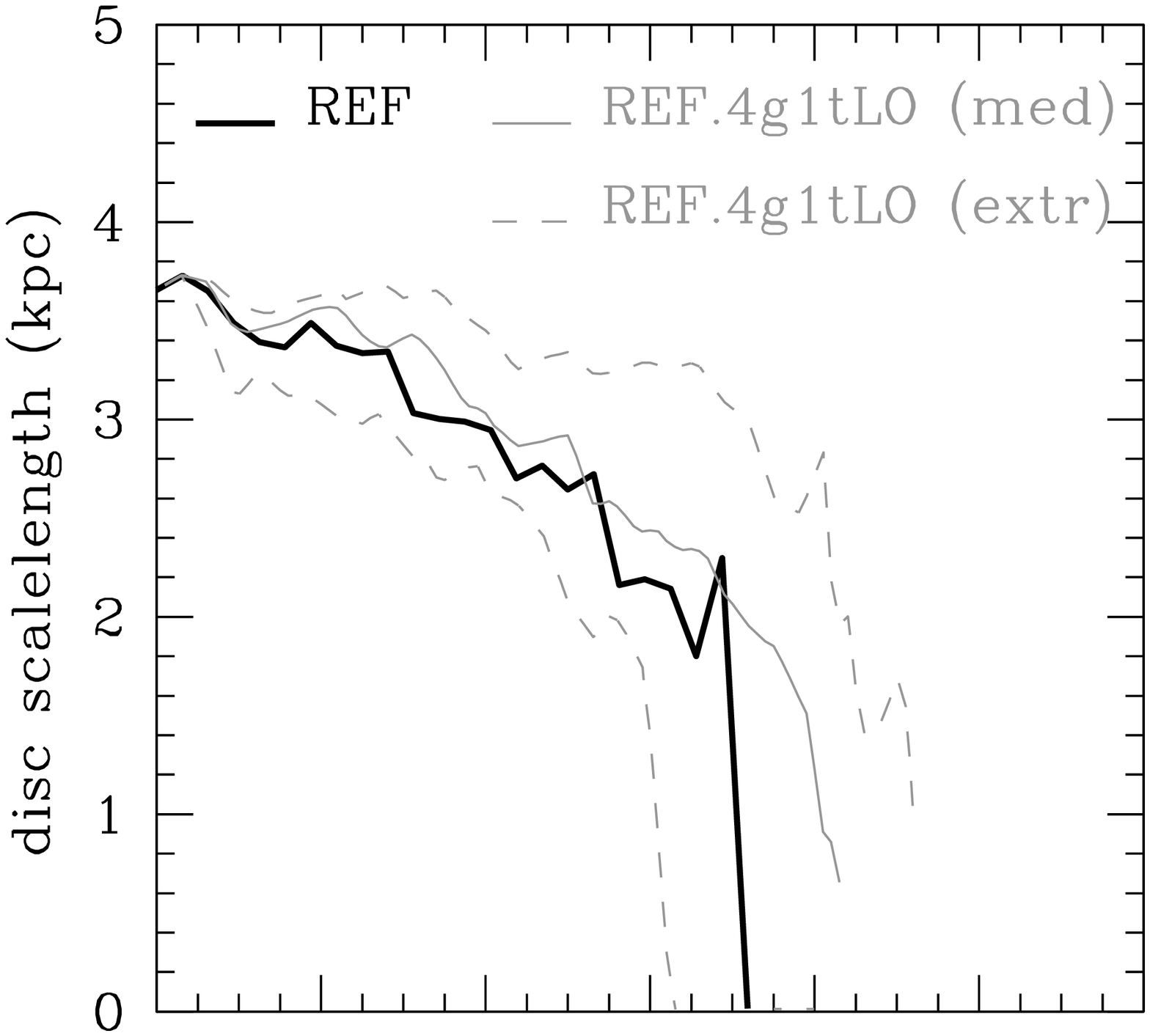}\hspace*{-11mm}
\includegraphics[width=65mm]{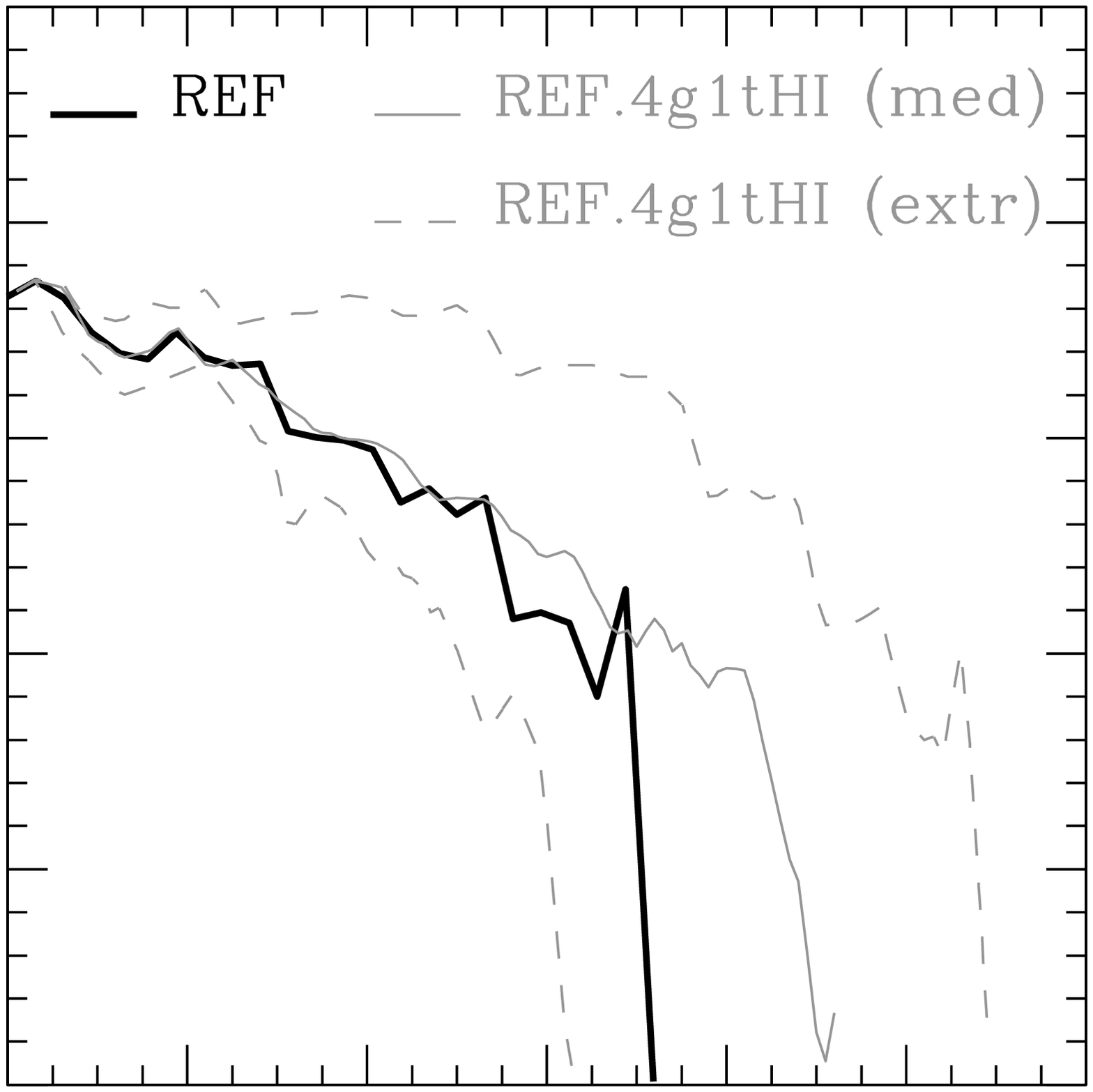}\hspace*{-11mm}
\includegraphics[width=65mm]{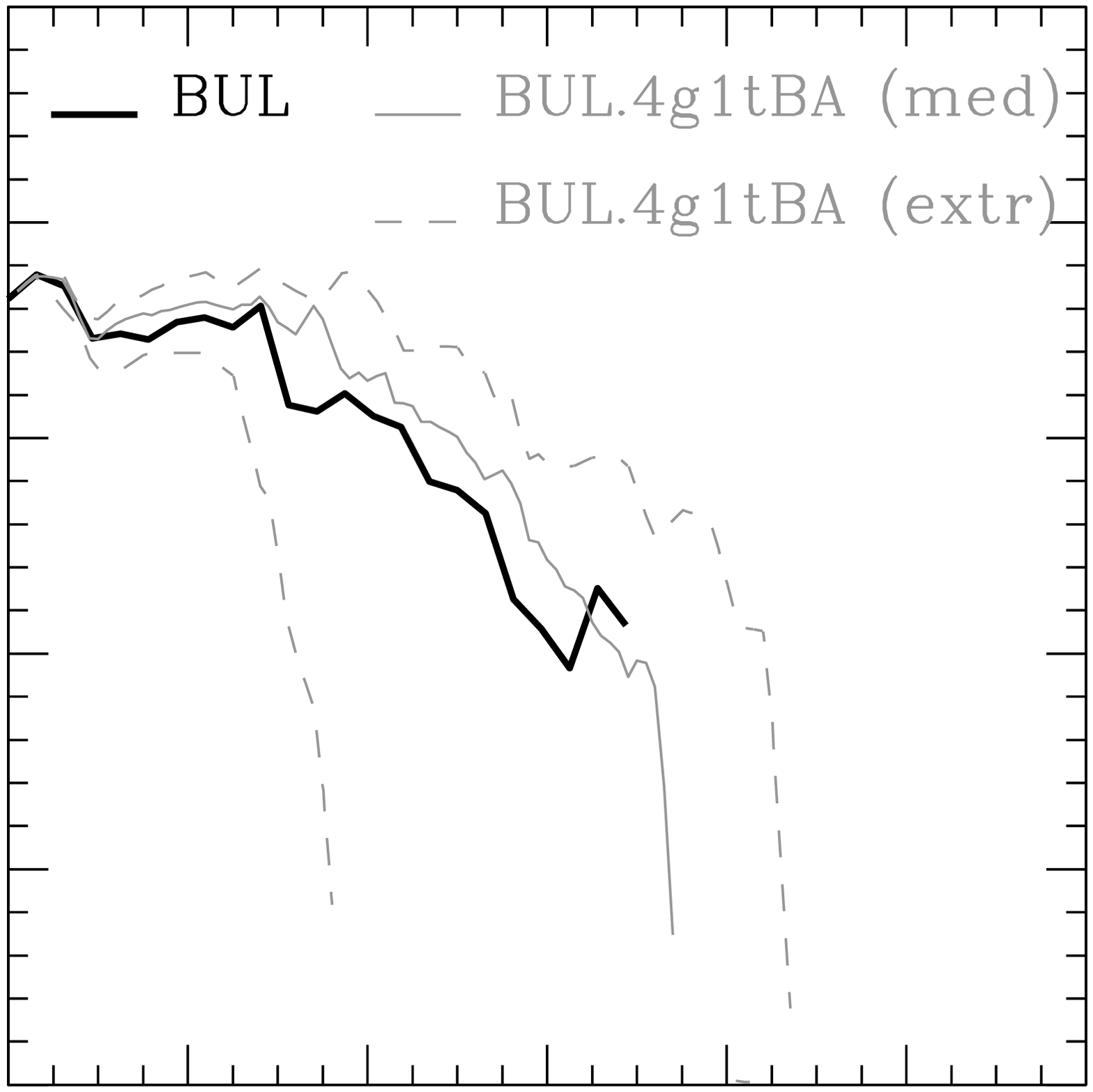}\vspace*{-11mm}\\
\includegraphics[width=65mm]{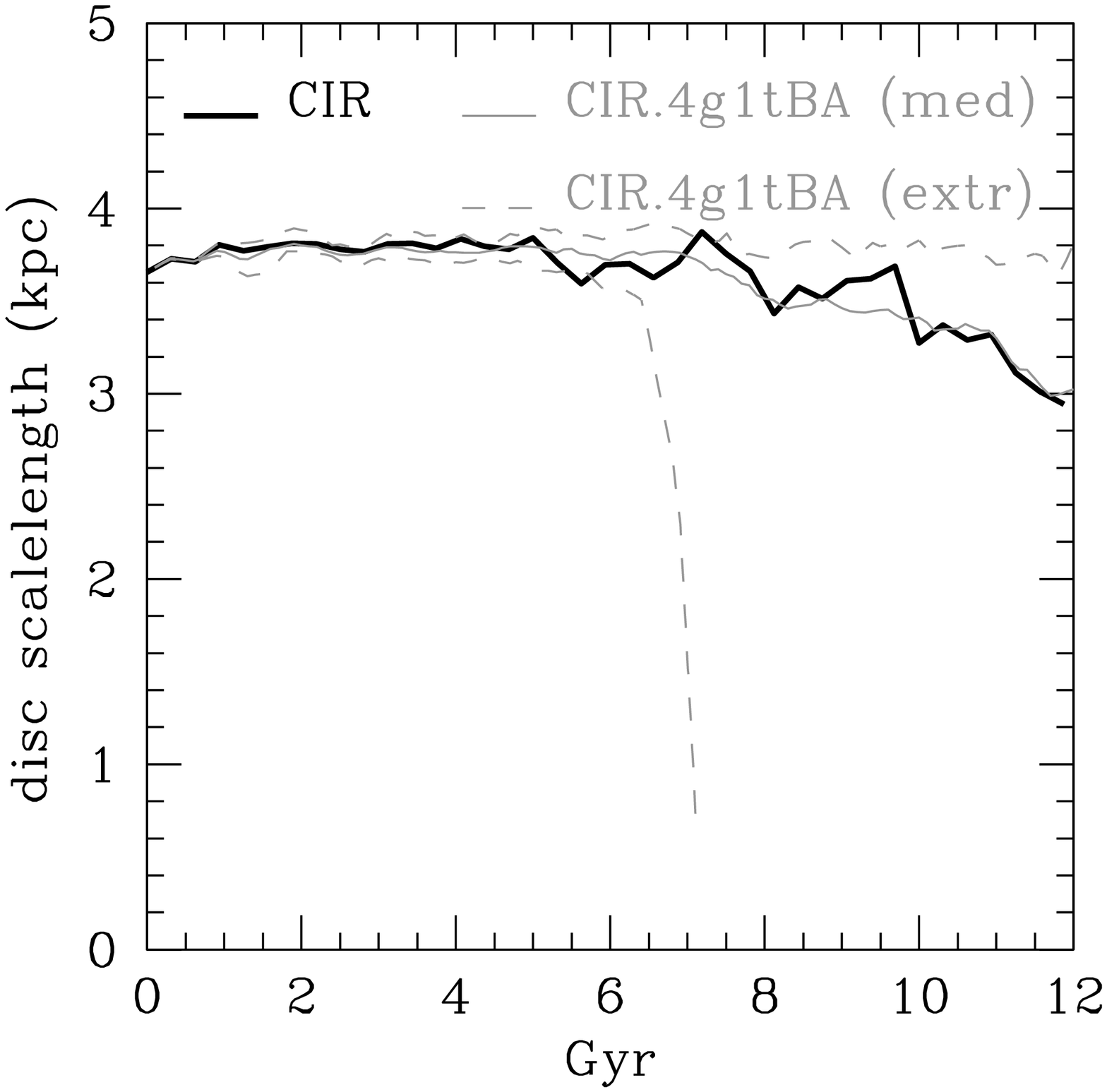}\hspace*{-11mm}
\includegraphics[width=65mm]{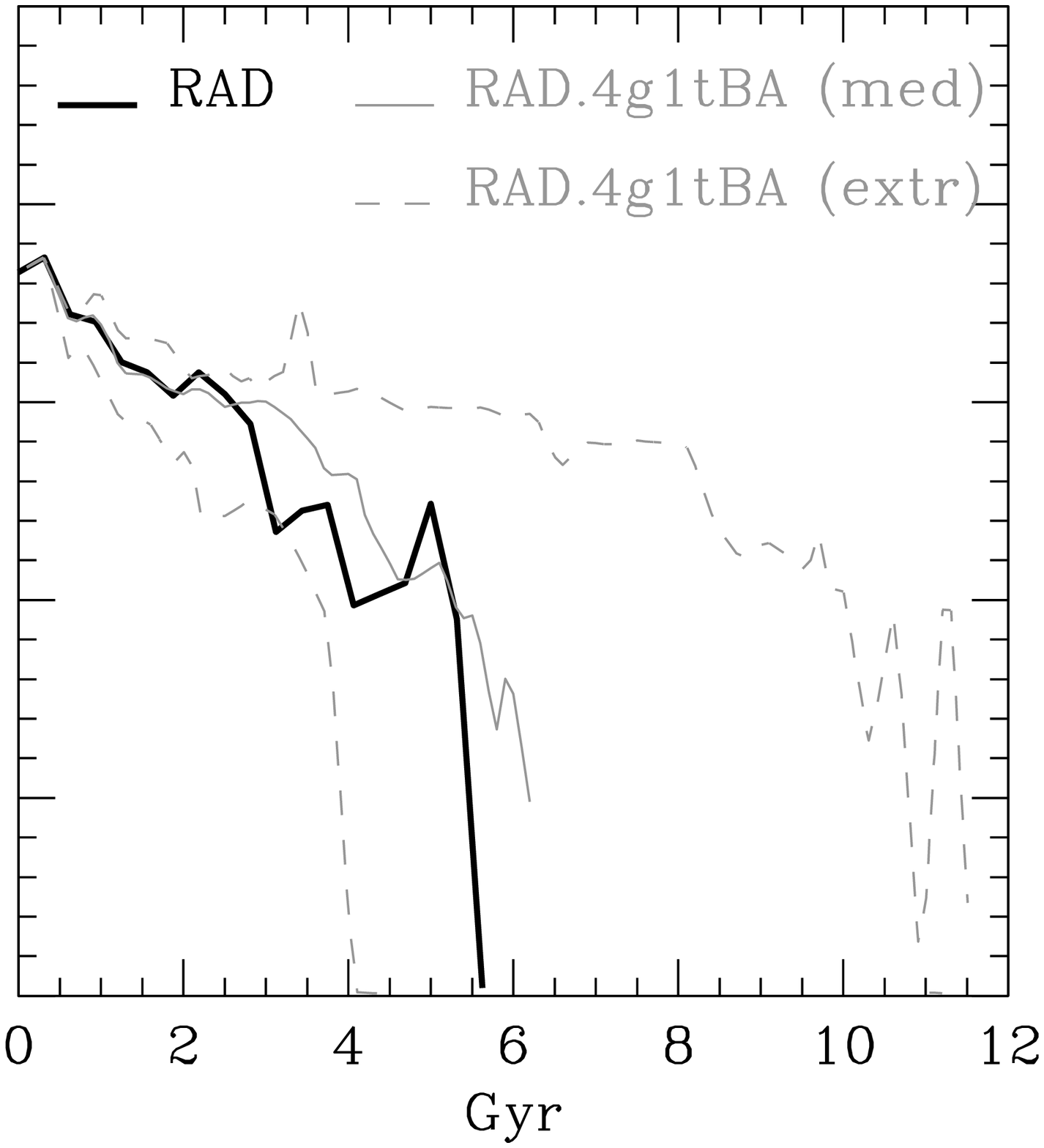}\hspace*{-11mm}
\includegraphics[width=65mm]{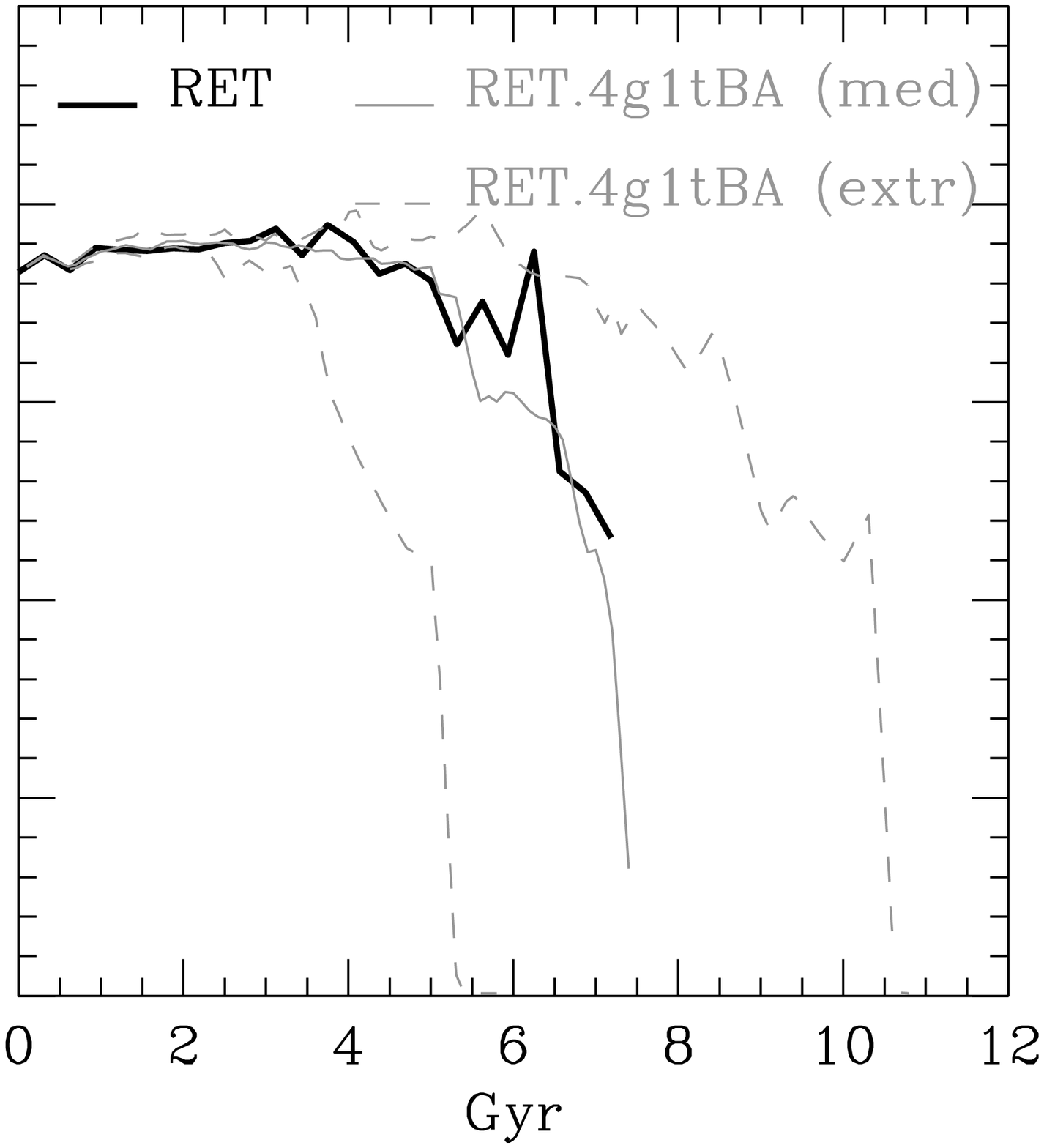}
\end{center}
\caption{Evolution of the scalelength of stellar discs (since infall until they 
are disrupted) in experiments covering different properties of the group 
population (number of members, combined total mass, mass distribution), disc's 
structure and (initial) orbital parameters of disc galaxies. Each panel compares 
the median (grey solid) and extreme cases (grey dashed) of the evolution due to 
the inclusion of group members to the corresponding case when no members are 
present in the group halo (black).
}
\label{scal-evol-disc}
\end{figure*}

\begin{figure*}
\begin{center}
\includegraphics[width=65mm]{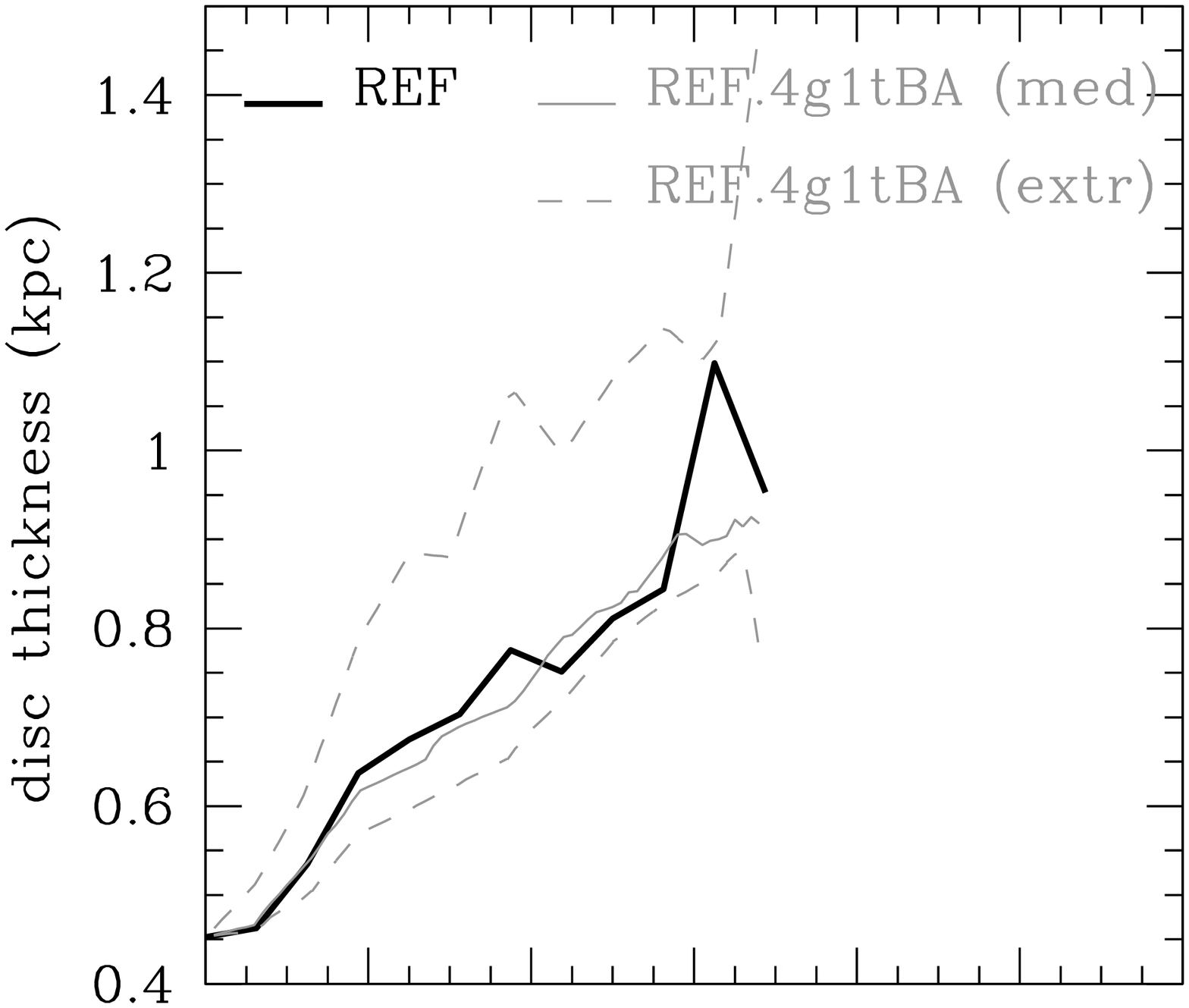}\hspace*{-11mm}
\includegraphics[width=65mm]{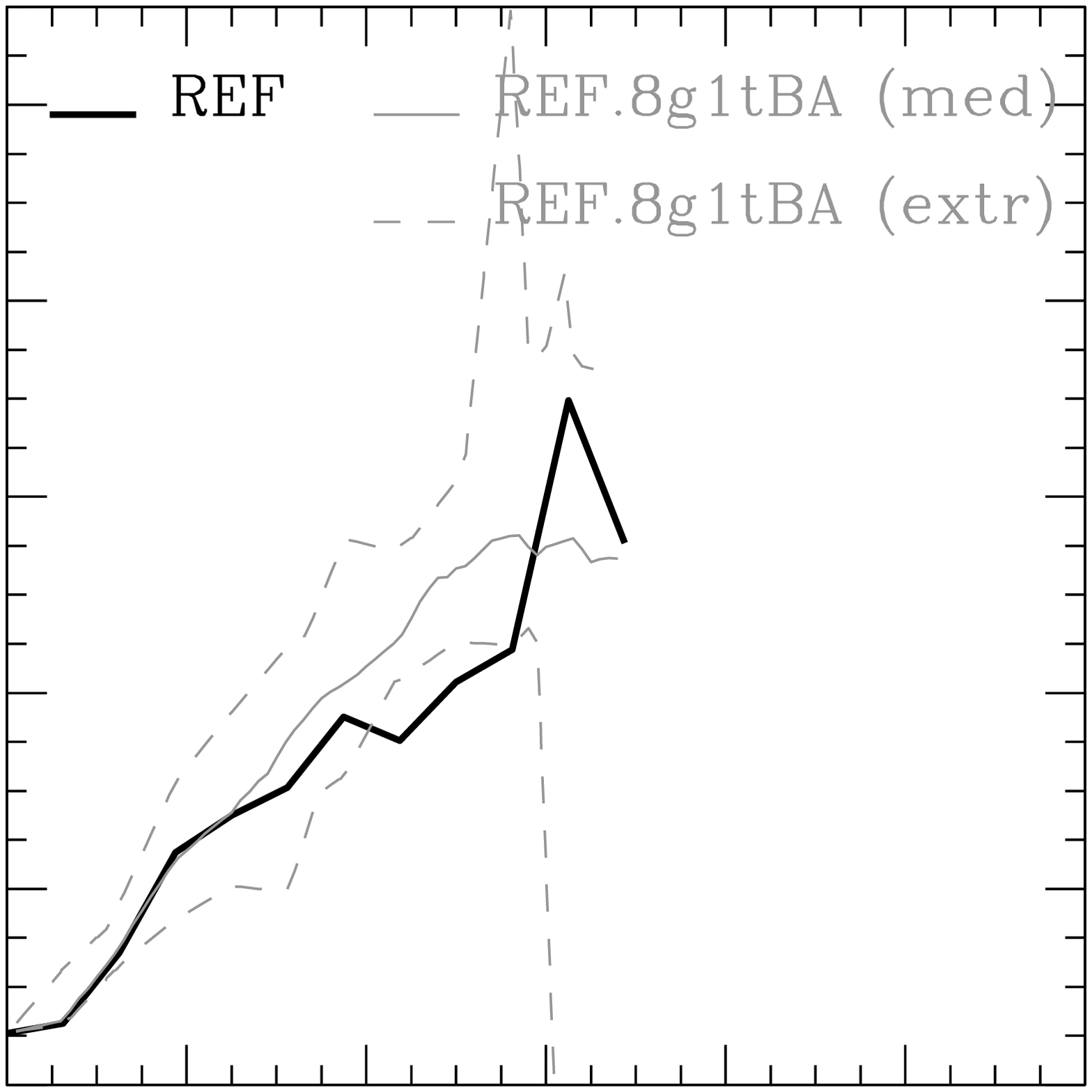}\hspace*{-11mm}
\includegraphics[width=65mm]{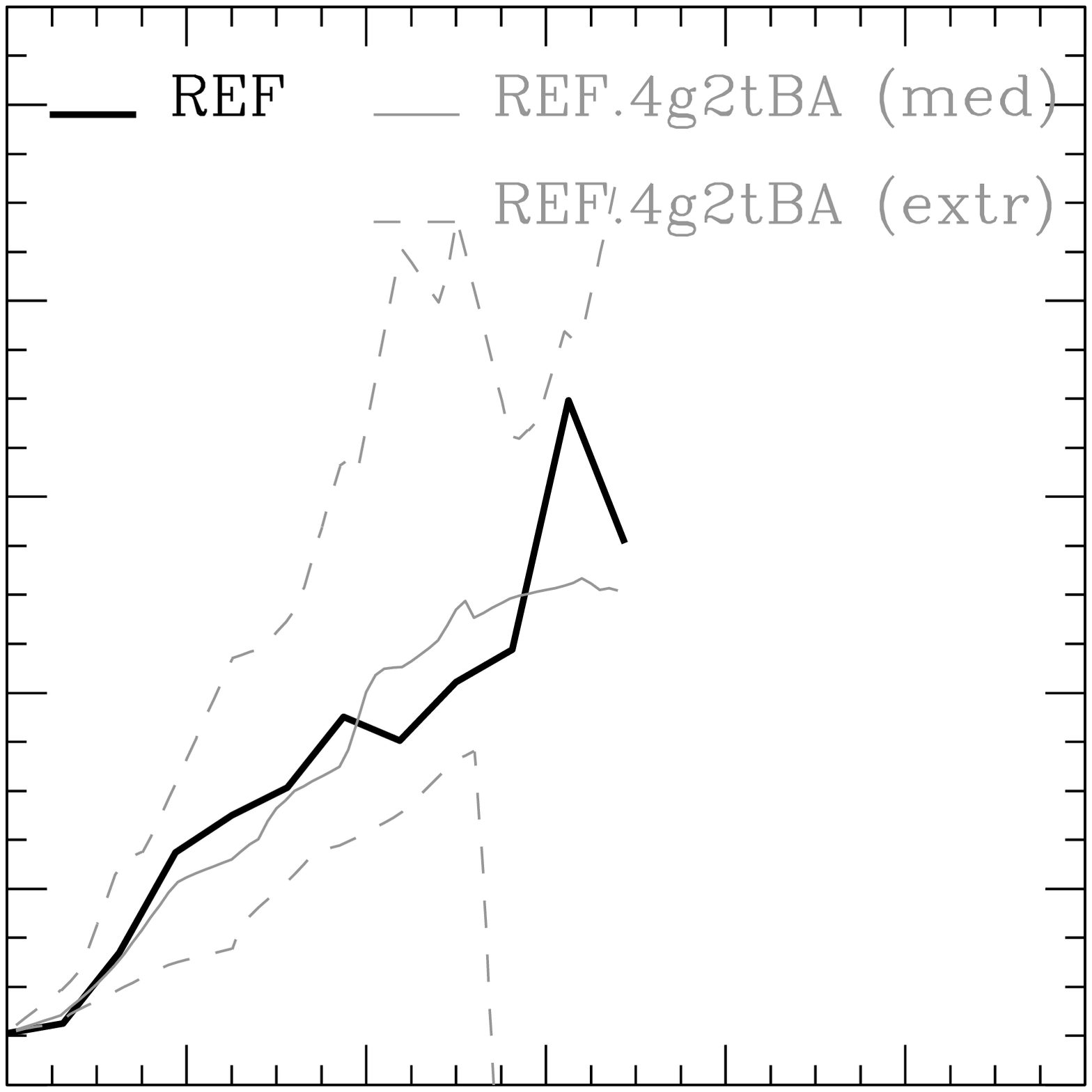}\vspace*{-11mm}\\
\includegraphics[width=65mm]{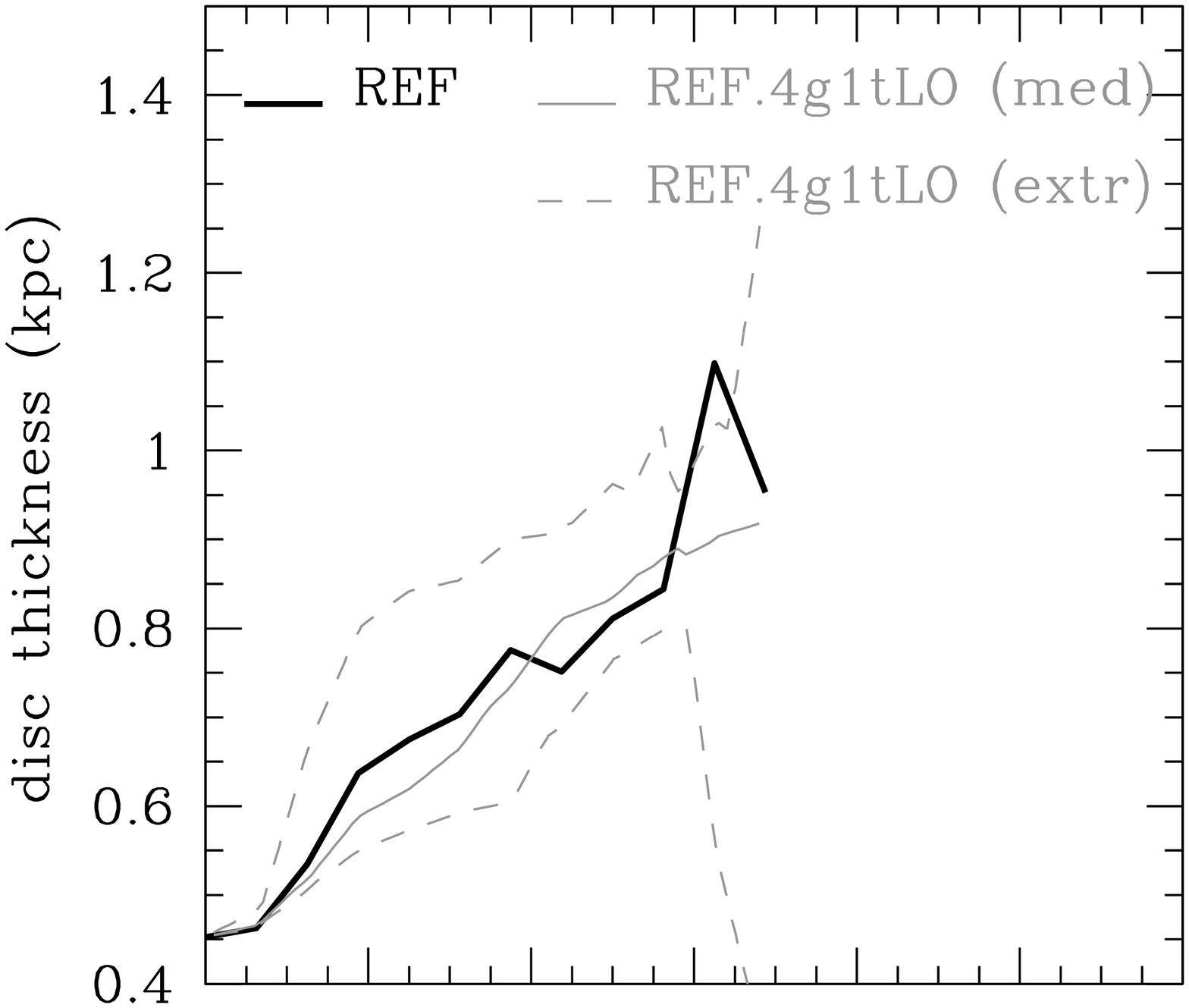}\hspace*{-11mm}
\includegraphics[width=65mm]{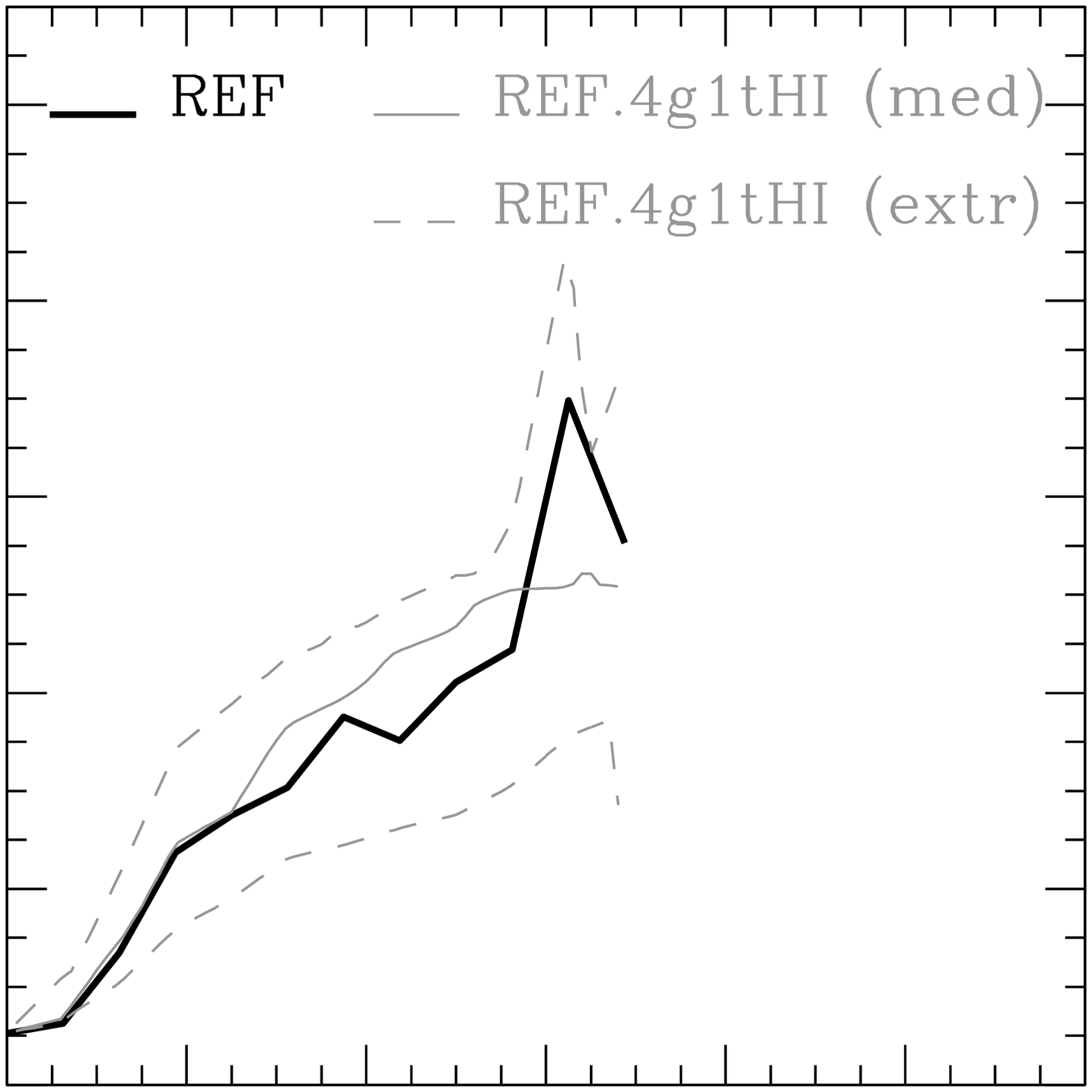}\hspace*{-11mm}
\includegraphics[width=65mm]{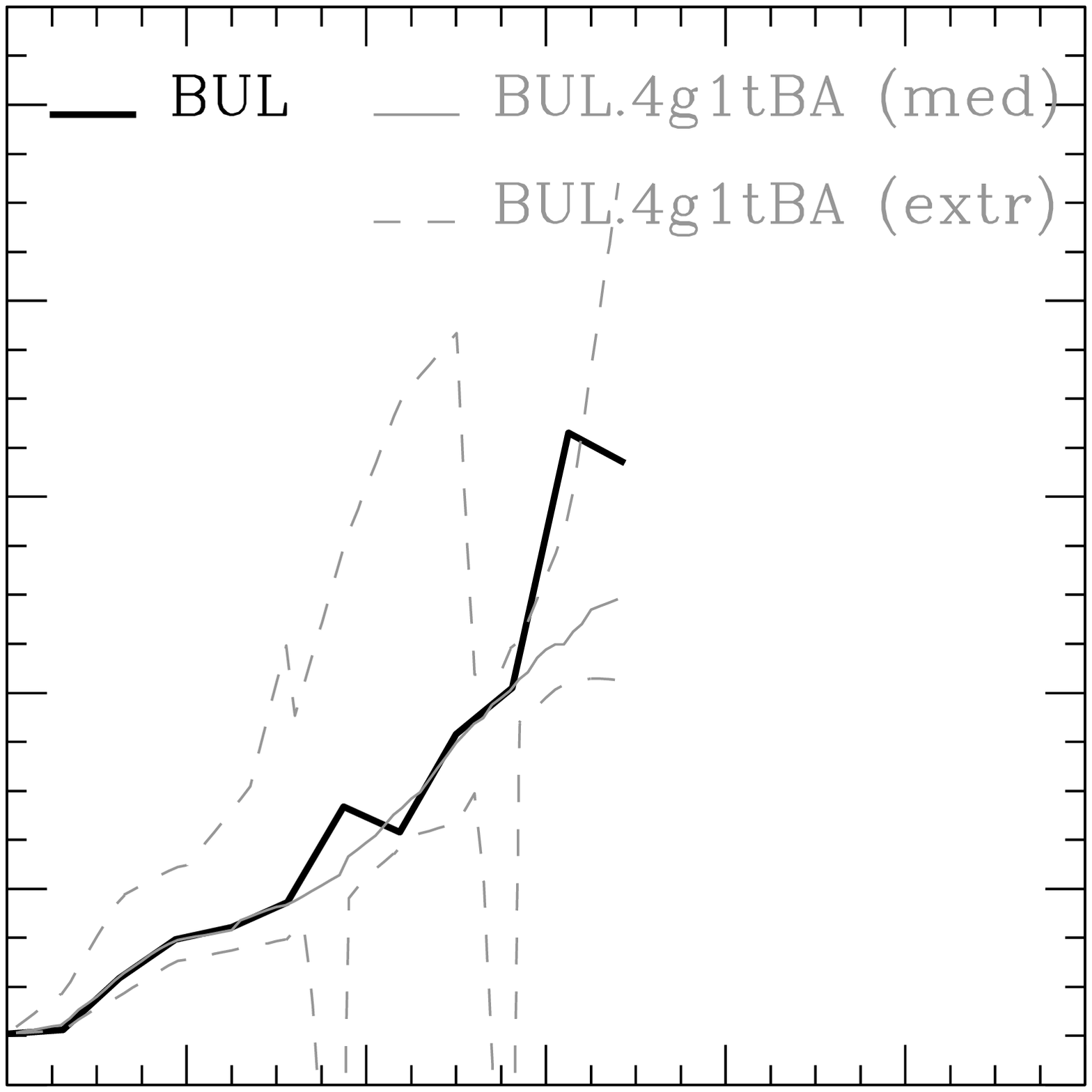}\vspace*{-11mm}\\
\includegraphics[width=65mm]{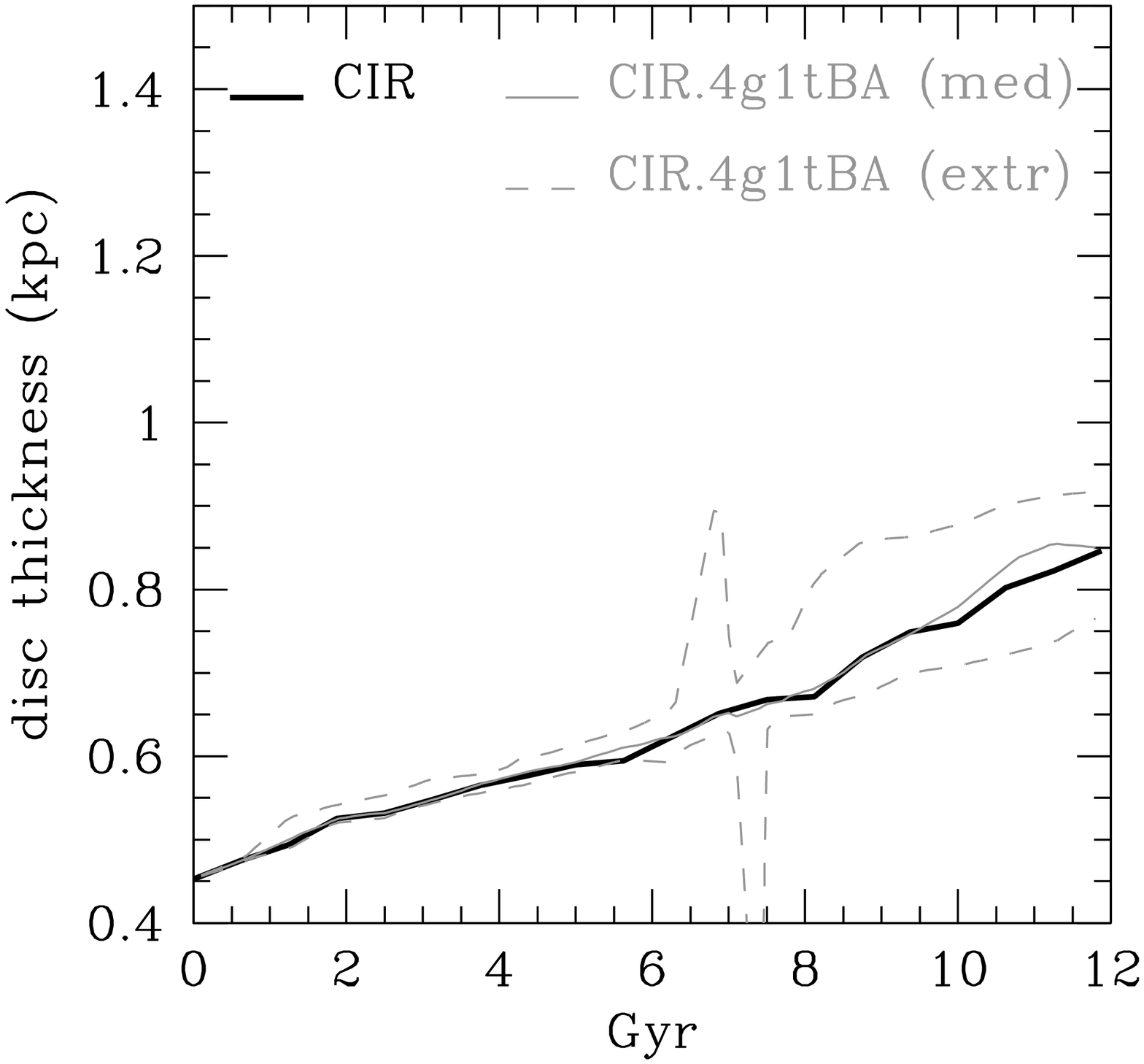}\hspace*{-11mm}
\includegraphics[width=65mm]{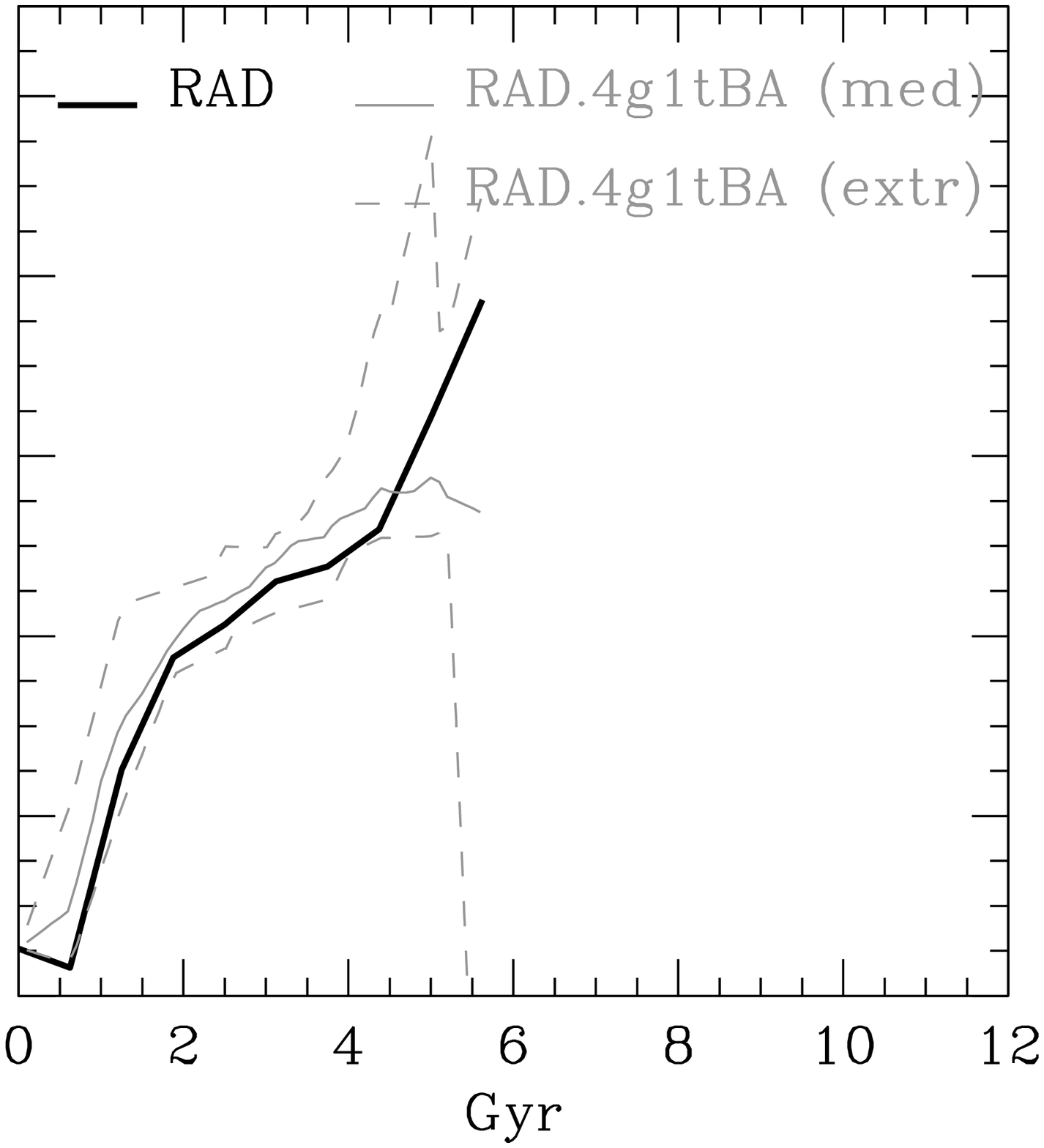}\hspace*{-11mm}
\includegraphics[width=65mm]{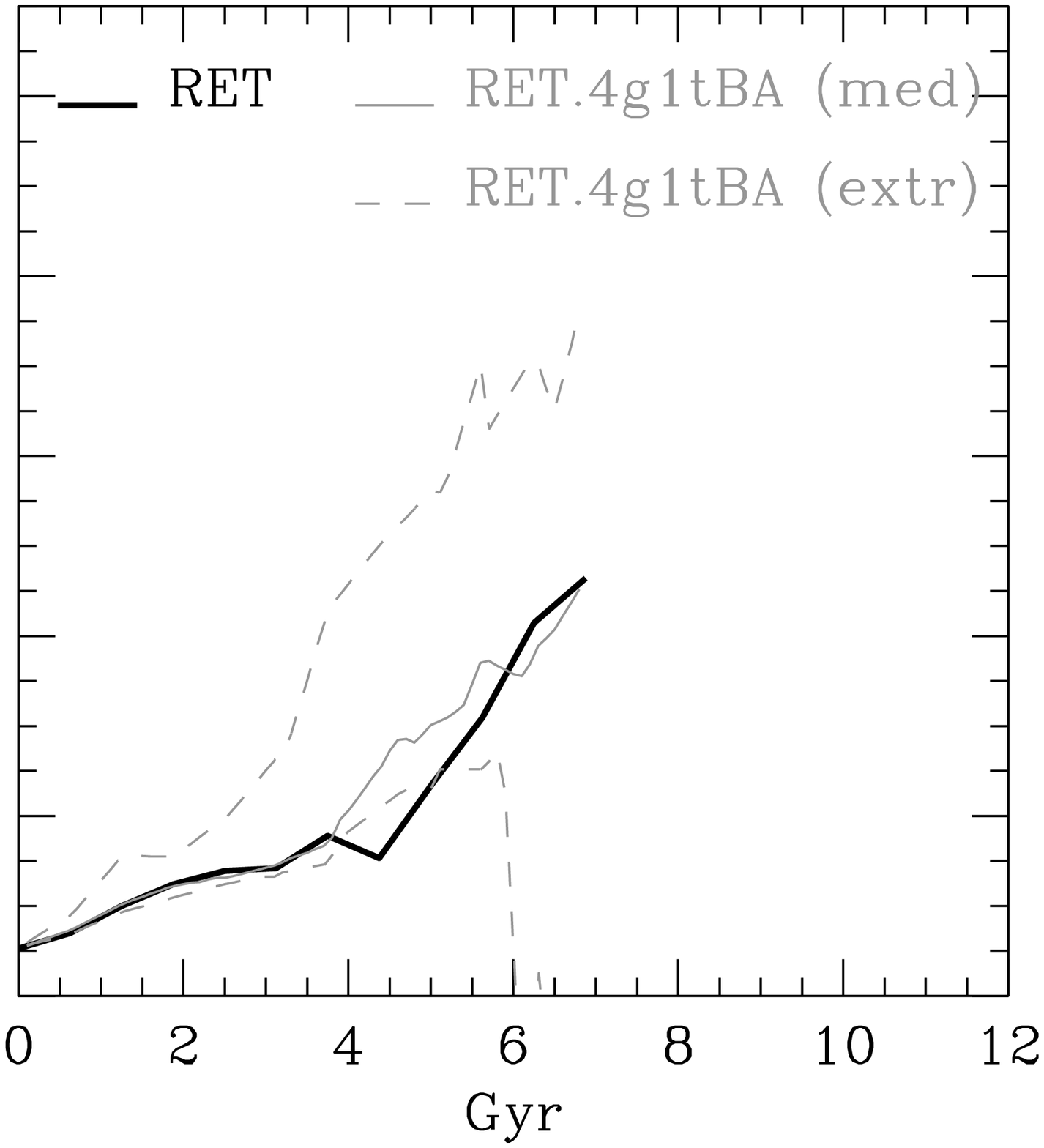}
\end{center}
\caption{Evolution of the (mean) thickness of stellar discs (since infall until 
they are disrupted) in experiments covering different properties of the group 
population (number of members, combined total mass, mass distribution), disc's 
structure and (initial) orbital parameters of disc galaxies. Each panel compares 
the median (grey solid) and extreme cases (grey dashed) of the evolution due to 
the inclusion of group members to the corresponding case when no members are 
present in the group halo (black).
}
\label{thick-evol-disc}
\end{figure*}

Figs.~\ref{scal-evol-disc} and \ref{thick-evol-disc} summarise the evolution of
the scalelength and mean thickness of disc galaxies in all our experiments. They
show the ``median'' evolution (over different initial orbital distributions of
group members), extreme cases, and the disc evolution in the corresponding
``control'' simulations where group members are not included. Even though the
inclusion of a population of group members introduces a significant ``scatter''
in the structural evolution of disc galaxies, in general the ``median'' evolution
of both the scalelength and mean thickness can be described remarkably well by
relatively simpler simulations that do not include group members. This is found
to be independent of the initial orbital eccentricity of the disc galaxy, its
initial inclination, the presence of a central bulge, the number of group
members, the total mass of the population of group members and their stellar mass
distribution. The ``scatter'' in the structural evolution of discs is driven
mainly by the total mass of the population of group members, and to a lesser
degree by the eccentricity of the orbits.

\bsp

\label{lastpage}

\end{document}